\documentclass[a4paper]{article}
\usepackage{geometry}
\usepackage{amsmath}
\usepackage{amssymb}
\usepackage{indentfirst}
\usepackage{mathdots}
\usepackage{cite}
\usepackage[colorlinks, linkcolor=blue, citecolor=blue]{hyperref}
\usepackage{xcolor}
\usepackage{mathrsfs}
\usepackage{booktabs}
\usepackage{enumitem}
\usepackage{authblk}
\usepackage{afterpage}
\usepackage{amsfonts,ntheorem}
\usepackage{threeparttable}
\usepackage{bm}
\usepackage{subfigure, epsfig}
\usepackage{setspace}
\usepackage{makecell}
\numberwithin{equation}{section}

\usepackage{algorithm}
\usepackage{algpseudocode}
\usepackage{comment}
\usepackage{cases}

\makeatletter

\newcommand{\Rmnum}[1]{\expandafter\@slowromancap\romannumeral #1@}
\makeatother

\newcommand\rank{\operatorname{rank}}
\newtheorem{theorem}{Theorem}[section]
\newtheorem{lemma}[theorem]{Lemma}
\newtheorem{corollary}[theorem]{Corollary}
\newtheorem{proposition}[theorem]{Proposition}

{ 
\theoremstyle{plain}   
\theorembodyfont{\normalfont} 

\newtheorem{definition}[theorem]{Definition}
\newtheorem{example}[theorem]{Example}
\newtheorem{remark}[theorem]{Remark}

}

\newenvironment{proof}{\noindent{\textbf{\emph{Proof.}}}}

\allowdisplaybreaks[4]

\definecolor{mydarkgreen}{RGB}{0,100,0}

\begin{document}

\title{\Large Criterion and Construction Framework for Impure Quantum $(r,\delta)$-LRCs from Optimal $(r,\delta)$-LRCs}
\author[1]{{\small{Kun Zhou}} \thanks{E-mail address: \texttt{kzhou@bimsa.cn}}}
\author[1]{{\small{Meng Cao}} \thanks{E-mail address: \texttt{mengcaomath@126.com} (corresponding author)}}
\affil[1]{\footnotesize{Beijing Institute of Mathematical Sciences and Applications (BIMSA), Beijing, 101408, China} }
\renewcommand*{\Affilfont}{\small\it}

\date{}
\maketitle

{\linespread{1.4}{

\vskip -10mm

\noindent {\small{{\bfseries{Abstract:}}
This paper aims to uncover new phenomena and provide new tools for quantum \((r,\delta)\)-LRCs derived from optimal \((r,\delta)\)-LRCs.
The initial motivation is to identify the main reason for the existence of impure cases in such quantum codes.
Our main result reveals that whether \(d = r+1\) is the key factor governing the impurity of quantum \((r,\delta)\)-LRCs.
We prove that, when \(d \neq r+1\), impure quantum codes derived from optimal \((r,\delta)\)-LRCs have bounded lengths. Conversely, when \(d = r+1\), we construct impure quantum \((r,\delta)\)-LRCs with unbounded lengths. The central tool developed in this work is a criterion for identifying impure quantum codes derived from optimal \((r,\delta)\)-LRCs. The main results include a criterion for impurity in quantum \((r,\delta)\)-LRCs (Theorem \ref{general-purity}), a construction framework for impure quantum \((r,\delta)\)-LRCs (Theorem \ref{impure-fram}), and two families of impure quantum \((r,\delta)\)-LRCs (Theorems \ref{impure-inf1} and \ref{impure-inf2}).
}}

\vspace{6pt}


\vspace{6pt}
}

\section{Introduction}\label{section1}

Classical locally recoverable codes (LRCs) \cite{Gopalan2012} and their generalization, \((r,\delta)\)-LRCs \cite{Prakash2012}, are essential in distributed storage systems. Their quantum analogues, quantum LRCs \cite{Golowich2025} and quantum \((r,\delta)\)-LRCs \cite{Galindo2024}, show potential applications in quantum data storage.

Classical \((r,\delta)\)-LRCs satisfy the Singleton-like bound
\begin{equation}
d \le n-k+1-\bigl(\lceil k/r\rceil-1\bigr)(\delta-1).
\label{eq:singleton-LRC}
\end{equation}
Codes attaining this bound are called optimal \((r,\delta)\)-LRCs, and they have been extensively studied (e.g., \cite{Huang2013,Kolosov2018}). Recently, quantum \((r,\delta)\)-LRCs have also received much attention (e.g., \cite{Matsumoto2026a}). Currently, effective methods to construct quantum \((r,\delta)\)-LRCs include the CSS construction and Hermitian construction \cite{Galindo2023,Matsumoto2026,CaoZhou2025-mp}. However, as noted in \cite{Galindo2026}, for quantum \((r,\delta)\)-LRCs with \(\delta\ge 3\), no unified tight bound is known for impure cases, unlike the case of \(\delta=2\), for which a bound on quantum \((r,2)\)-LRCs with parameters $[[n,\kappa,d]]_q$ was given in \cite{Golowich2025}:
\begin{equation}
  \kappa \;\le\; n-2(d-1)
  -\left\lfloor \frac{n-(d-1)}{r+1}\right\rfloor
  -\left\lfloor \frac{n-2(d-1)-\left\lfloor\frac{n-(d-1)}{r+1}\right\rfloor}{r+1}.\right\rfloor
  \label{eq:delta2-bound}
\end{equation}
Impure codes derived from the CSS construction or Hermitian construction can exhibit essential differences; for instance, they may violate different bounds compared to
their pure counterparts \cite{Galindo2026}. Thus, studying the impurity is unavoidable for quantum \((r,\delta)\)-LRCs generated by these constructions.

Previous works on impurity criteria focused on specific families of Euclidean or Hermitian dual-containing \((r,\delta)\)-LRCs \cite{Galindo2023,Galindo2026,CaoZhou2025-mp}. For example, the purity was studied for quantum codes from MP codes \cite{Galindo2023} and from \(J\)-affine variety codes \cite{Galindo2026}. In contrast, we explore the impurity for quantum codes induced from general optimal \((r,\delta)\)-LRCs. We derive necessary conditions depending only on the parameters $r$, $\delta$, $k$, and $d$. The result shows that once these parameters are fixed, then if \(d\neq r+1\), impurity can only occur for bounded code lengths. The main tool is the decomposition theorem proposed in \cite[Theorem 3.6]{ZhouCao2025-Opt}.

On the other hand, the construction of impure quantum \((r,\delta)\)-LRCs from the CSS construction or Hermitian construction is still in its early stage. Recently, Galindo et al. \cite{Galindo2026} used the CSS construction to obtain infinite families of impure quantum \((r,\delta)\)-LRCs that violate various pure bounds. A main difficulty is determining the minimum distance of quantum codes, especially for those derived from optimal \((r,\delta)\)-LRCs. Guided by Theorem \ref{general-purity}, we aim to establish a general construction framework for impure quantum \((r,\delta)\)-LRCs. As concrete applications, we construct two families of impure quantum \((r,\delta)\)-LRCs via the CSS construction and Hermitian construction, respectively, showing that whether \(d = r+1\) is indeed a key factor controlling the purity.

Our main results include a criterion for impure quantum $(r,\delta)$-LRCs derived from optimal $(r,\delta)$-LRCs:
\begin{theorem}\emph{(see also Theorem \ref{general-purity})}
Let $\mathcal{C}$ be an optimal $(r,\delta)$-LRC with parameters $[n,k,d]_q$ (resp. $[n,k,d]_{q^2}$). Suppose $\mathcal{C}$ is Euclidean or Hermitian dual-containing.
If $\mathcal{C}$ induces an impure quantum code, then one of the following conditions must hold:
\begin{itemize}
\item[(i)] $d=r+1$;

\item[(ii)] $d\neq r+1$ and $\lceil\frac{k}{r}\rceil\leq \max\{\frac{2(r-1)}{r+1-d}, \frac{2(r-1)}{r+\delta-d}+1\}$.
\end{itemize}
\end{theorem}

Furthermore, we establish Theorem \ref{impure-iff} as a main tool for constructing impure quantum \((r,\delta)\)-LRCs. As an application, we present the following construction, which corresponds to condition (i) of Theorem \ref{general-purity}.
\begin{theorem}\emph{(see also Theorem \ref{impure-fram})}
Given integers $r,\delta\geq 2$ satisfying $\delta< r+1< 2\delta$. Suppose $\mathcal{C}$ has a parity-check matrix $H$ of the form \eqref{pari-gen-z11} satisfying (i)-(iii) of Theorem \ref{impure-fram}. Then, the code $\mathcal{C}$ has parameters $[t(r+\delta-1),tr-r-1+\delta,r+1]_q$ (resp. $[t(r+\delta-1),tr-r-1+\delta,r+1]_{q^2}$), and it induces an impure quantum \((r,\delta)\)-LRC with parameters
\begin{align}
[[t(r+\delta-1),(t-2)(r+1-\delta), \geq 2\delta]]_{q}. \label{impure-quan-y1}
\end{align}
\end{theorem}
Moreover, using Corollary \ref{impure-fram-v2}, we obtain concrete impure quantum $(r,\delta)$-LRCs via condition (ii) of Theorem \ref{general-purity}.

Combining these results, we observe the following phenomena:
\begin{itemize}
  \item If $d=r+1$, there exist infinite families of optimal $(r,\delta)$-LRCs of unbounded lengths whose induced quantum codes are impure (see Theorem \ref{impure-inf1} and Theorem \ref{impure-inf2}).
  \item If $d\neq r+1$, there is no infinite family of optimal $(r,\delta)$-LRCs of unbounded lengths whose induced quantum codes are impure (see Theorem \ref{general-purity} and Remark \ref{purity-finite}).
\end{itemize}

The paper is organized as follows. Section~2 recalls preliminaries, including optimal \((r,\delta)\)-LRCs, quantum \((r,\delta)\)-LRCs, and the decomposition theorem. Section~3 presents the impurity criterion. Section~4 provides a construction framework for impure quantum $(r,\delta)$-LRCs, and uses the CSS construction and Hermitian construction to obtain two families of impure quantum \((r,\delta)\)-LRCs derived from optimal ones.

\section{Preliminaries}\label{sec:pre}

\subsection{Notation and terminology}
For a matrix $A=(a_{i,j})$ over $\mathbb{F}_{q}$, its {\it transpose} is denoted by $A^{\top}:=(a_{j,i})$.
Let $\mathbb{N}=\{0,1,2,\ldots\}$ denote the set of natural numbers, and let $\mathbb{N}^+=\mathbb{N} \setminus \{0\}$ denote the set of positive integers.
For $n\in \mathbb{N}^+$, define the integer interval $[n]:=\{1,2,\ldots,n\}$.
For a finite set $S$, let $|S|$ denote its {\it cardinality}.
For a set of vectors $T$, let $\langle T\rangle$ denote the {\it linear span} of $T$.

\subsection{Linear codes}
Denote by $[n,k,d]_{q}$ a {\it linear code} over $\mathbb{F}_{q}$ of length $n$, dimension $k$ and minimum (Hamming) distance. It is a $k$-dimensional subspace of the vector space $\mathbb{F}_{q}^{n}$. By the {\it Singleton bound}, the minimum distance satisfies $d\leq n-k+1$. If the equality holds, the code is called a {\it maximum distance separable (MDS) code}.

For a linear code $\mathcal{C}$ of length $n$ over $\mathbb{F}_{q}$ (resp. $\mathbb{F}_{q^{2}}$), its {\it Euclidean dual} $\mathcal{C}^{\perp_{\mathrm{E}}}$ (resp. {\it Hermitian dual} $\mathcal{C}^{\perp_{\mathrm{H}}}$) are respectively defined by
\begin{align*}
\mathcal{C}^{\perp_{\mathrm{E}}}=\Big\{\mathbf{x}=(x_{1},\ldots,x_{n})\in\mathbb{F}_{q}^{n}:\
\langle\mathbf{x},\mathbf{y}\rangle=\sum_{i=1}^n x_i y_i=0 \ \mathrm{for} \ \mathrm{all} \  \mathbf{y}=
(y_{1},\ldots,y_{n})\in \mathcal{C}\Big\}
\end{align*}
and
\begin{align*}
\mathcal{C}^{\perp_{\mathrm{H}}}=\Big\{\mathbf{x}=(x_{1},\ldots,x_{n})\in\mathbb{F}_{q^{2}}^{n}:\
\langle\mathbf{x},\mathbf{y}\rangle_{\mathrm{H}}=\sum_{i=1}^n x_i y_i^{q}=0 \ \mathrm{for} \ \mathrm{all} \  \mathbf{y}=
(y_{1},\ldots,y_{n})\in \mathcal{C}\Big\}.
\end{align*}
If $\mathcal{C}\subseteq\mathcal{C}^{\perp_{\mathrm{E}}}$ (resp. $\mathcal{C}\subseteq\mathcal{C}^{\perp_{\mathrm{H}}}$), then $\mathcal{C}$ is said to be {\it Euclidean self-orthogonal} (resp. {\it Hermitian self-orthogonal}).
If $\mathcal{C}^{\perp_{\mathrm{E}}}\subseteq\mathcal{C}$ (resp. $\mathcal{C}^{\perp_{\mathrm{H}}}\subseteq\mathcal{C}$), then $\mathcal{C}$ is said to be
{\it Euclidean dual-containing} (resp. {\it Hermitian dual-containing}).

\subsection{Optimal $(r,\delta)$-LRCs}\label{LRC}

For an $[n, k, d]_q$ linear code $\mathcal{C}$, the $i$th symbol $c_i$, where $i\in[n]$,
of $\mathcal{C}$ is said to have {\it $(r, \delta)$-locality} if there exists a subset
$S_i$ $\subseteq [n]$ containing $i$ and a punctured code
$\mathcal{C}|_{S_i}$ such that the length $|S_i| \leq r + \delta -1$ and the minimum distance $d(\mathcal{C}|_{S_i})\geq \delta $, where $\mathcal{C}|_{S_i}$ is the code $\mathcal{C}$ punctured on the coordinate set $[n]\backslash S_i$ by deleting the components indexed by the set $[n]\backslash S_i$ in each codeword of $\mathcal{C}$. The code $\mathcal{C}$ is said to be an {\it $(r, \delta)$-LRC} if every symbol has $(r,\delta)$-locality.

Suppose that $H = (\mathbf{h}_1, \mathbf{h}_2, \ldots , \mathbf{h}_n)$ is an $m\times n$ matrix, where $\mathbf{h}_i$ is a column vector of length $m$. The {\it support} of $H$ is defined as
$$\text{Supp}(H) :=\{i \in [n]:\; \mathbf{h}_i \neq \mathbf{0}\},$$
where $\mathbf{0}$ denotes the zero column vector of length $m$.

\begin{lemma}(\!\!\cite[Lemma 2]{Luo2023Three}) \label{local}
Let $\mathcal{C}$ be an $[n, k, d]_q$ linear code. Let $r$ and $\delta$ be positive integers, where $\delta > 1$. The $\alpha$th code symbol of $\mathcal{C}$ has $(r,\delta)$-locality if and only if there exists an $m\times n$ matrix $H$ over $\mathbb{F}_q$ with the following properties:
\begin{itemize}
\item [(1)] $\alpha\in\text{Supp}(H)$;

\item [(2)] $|\text{Supp}(H)|\leq  r +\delta-1$ and any $\delta-1$ nonzero
columns of $H$ are linearly independent over $\mathbb{F}_q$;

\item [(3)] $H\mathbf{c}^{\top} = \mathbf{0}$ for every codeword $\mathbf{c}\in \mathcal{C}$.
\end{itemize}
\end{lemma}

An $(r, \delta)$-LRC always satisfy the Singleton-like bound \eqref{eq:singleton-LRC}. It is called \emph{optimal} when it touches this bound.

\subsection{Quantum codes}
A $q$-ary quantum code of length $n$, dimension $q^{k}$, and minimum distance $d$ is denoted by $[[n,k,d]]_{q}$. It is a $q^{k}$-dimensional subspace of the $q^{n}$-dimensional complex Hilbert space $(\mathbb{C}^{q})^{\otimes n} \cong \mathbb{C}^{q^{n}}$. This quantum code is capable of detecting up to $d-1$ quantum errors and correcting up to $\left\lfloor \frac{d-1}{2} \right\rfloor$ quantum errors. The minimum distance $d$ of an $[[n,k,d]]_{q}$ quantum code is controlled by the \emph{quantum Singleton bound}, which states that $2d \leq n-k+2$. A quantum code meeting this bound with equality, i.e., $2d = n - k + 2$, is called a
\emph{quantum MDS code}. For further background on quantum codes, we refer the reader to \cite{Ashikhmin2001Nonbinary,Calderbank1998Quantum,Ketkar2006Nonbinary,Knill1997Theory,Gottesman1997Stabilizer,Grassl2004On,Rains1999Nonbinary}.

The famous CSS construction, named after Calderbank and Shor \cite{Calderbank1996Good} and Steane \cite{Steane1996Simple}, provides an efficient method for yielding quantum codes from a pair of linear codes.

\begin{proposition}{\rm (\!\!\cite{Calderbank1996Good,Steane1996Simple}, CSS Construction)} \label{css}
Let $\mathcal{C}_{i}$ be an $[n,k_{i},d_{i}]_{q}$ linear code for $i=1,2$ such that $\mathcal{C}_{2}^{\bot_{\mathrm{E}}}\subseteq\mathcal{C}_{1}$.
Then, there exists a quantum code with parameters $[[n,k_{1}+k_{2}-n,d]]_{q}$, where
\begin{align}\label{distanceee}
d&=\begin{cases}
\mathrm{min}\{d_1,d_2\}, & \mathrm{if}\ \mathcal{C}_{2}^{\bot_{\mathrm{E}}}=\mathcal{C}_{1},\\
\mathrm{min}\{\mathrm{wt}(\mathcal{C}_{1}\backslash\mathcal{C}_{2}^{\bot_{\mathrm{E}}}),\mathrm{wt}(\mathcal{C}_{2}\backslash\mathcal{C}_{1}^{\bot_{\mathrm{E}}})\}, & \mathrm{if}\ \mathcal{C}_{2}^{\bot_{\mathrm{E}}}\neq\mathcal{C}_{1}.
\end{cases}
\end{align}
\end{proposition}

The Hermitian construction, shown in the following proposition, gives an effective method for producing quantum codes from Hermitian dual-containing codes.

\begin{proposition}{\rm (\!\!\cite{Ashikhmin2001Nonbinary}, Hermitian construction)}\label{proposition6}
Let $\mathcal{C}$ be an $[n,k,d]_{q^{2}}$ Hermitian dual-containing code. Then, there exists a quantum code with parameters $[[n,k_{1}+k_{2}-n,d_\text{H}]]_{q}$, where
\begin{align}\label{distanceee2}
d_\text{H}&=\begin{cases}
d, & \mathrm{if}\ \mathcal{C}^{\bot_{\mathrm{H}}}=\mathcal{C},\\
\mathrm{wt}(\mathcal{C}\backslash\mathcal{C}^{\bot_{\mathrm{H}}}), & \mathrm{if}\ \mathcal{C}^{\bot_{\mathrm{H}}}\neq\mathcal{C}.
\end{cases}
\end{align}
\end{proposition}

\subsection{Quantum $(r, \delta)$-LRCs via the CSS construction and Hermitian construction}

In \cite{Galindo2024}, Galindo et al. established a connection between Hermitian dual-containing (resp. Euclidean dual-containing) classical codes and quantum $(r, \delta)$-LRCs. Building upon this result, they further investigated optimal pure quantum $(r, \delta)$-LRCs constructed from classical codes and introduced the following definition.

Let $\mathcal{C}$ be a \(q^2\)-ary (resp. \(q\)-ary) linear code. Assume that $\mathcal{C}$ is Hermitian dual-containing (resp. Euclidean dual-containing). Then, the quantum code \(Q(\mathcal{C})\) induced by $\mathcal{C}$ is a quantum \((r,\delta)\)-LRC if and only if $\mathcal{C}$ is a classical \((r,\delta)\)-LRC (see \cite[Theorem 28]{Galindo2024}). Moreover, the authors introduced the following definition.

\begin{definition}{\rm (\!\!\cite[Definition 33]{Galindo2024})}\label{quan-LRC}
Let $\mathcal{C}\subseteq \mathbb{F}_{q^2}^n$ (resp. $\mathcal{C}\subseteq \mathbb{F}_{q}^n$) be a linear code. Suppose $\mathcal{C}$ is Hermitian dual-containing (resp. Euclidean dual-containing), $\dim(\mathcal{C})=\frac{n+k}{2}$ and $\mathcal{C}$ is an $(r, \delta)$-LRC. If $\mathcal{C}$ is pure and if the induced quantum code with parameters $[[n,k,d(\mathcal{C})]]_{q}$ satisfies
\begin{align}
k + 2d(\mathcal{C}) + 2\bigg(\bigg\lceil \frac{n+k}{2r}\bigg\rceil-1\bigg)(\delta-1)=n+2,\label{quan-inqu}
\end{align}
then the induced quantum code with parameters $[[n,k, d(\mathcal{C})]]_{q}$ is said to be an optimal pure quantum $(r, \delta)$-LRC.
\end{definition}

\begin{remark}\label{optimal}
Since $k=2\dim(\mathcal{C})-n$ in Definition \ref{quan-LRC}, we can verify that Eq. \eqref{quan-inqu} is equivalent to
\begin{align*}
d(\mathcal{C})=n-\dim(\mathcal{C})+1-\bigg(\bigg\lceil \frac{\dim(\mathcal{C})}{r}\bigg\rceil-1\bigg)(\delta-1).
\end{align*}
This is equivalent to stating that $\mathcal{C}$ is an optimal classical $(r,\delta)$-LRC. Consequently, within the framework of Definition \ref{quan-LRC}, an Hermitian
dual-containing (resp. Euclidean dual-containing) $(r,\delta)$-LRC is optimal if and only if its induced pure quantum $(r,\delta)$-LRC is optimal.
\end{remark}

\subsection{Decomposition theorem}
Now, we recall the result in \cite[Theorem 3.4]{ZhouCao2025-Opt}. Let $\mathcal{C}$ be an $[n, k, d]_q$ linear code with generator matrix $G=(\mathbf{c}_1,\ldots,\mathbf{c}_n)$. Recall the following canonical set given in \cite{ZhouCao2025-Opt}:
\begin{align}
C=\{(1,\mathbf{c}_1),\ldots,(n,\mathbf{c}_n)\}\subseteq [n]\times \mathbb{F}_{q}^k. \label{setC}
\end{align}

\begin{theorem}(\emph{Decomposition Theorem})\label{Stru}
Suppose $\mathcal{C}$ is an optimal $(r,\delta)$-LRC with parameters $[n,k,d]_q$. Let $C$ be the set defined as in Eq. \eqref{setC}. Then, $C$ must belong to one of following two decompositions:\\
\underline{Case (I)}:
$$C=C_1\cup...\cup C_{t-1}\cup \{(j_1,\mathbf{c}_{j_1}),...,(j_i,\mathbf{c}_{j_{s_t}})\}\cup U,$$
where $t=\lceil \frac{k}{r}\rceil,\;C_i\subseteq C$ and $\{j_1,...,j_{s_t}\}\subseteq [n]$, and $\{j_1,...,j_{s_t}\}$ is contained in some $T\subseteq C$ such that $\mathcal{C}|_{T}$ is a local protection code, and
\begin{itemize}
\item [(1)] Each $\mathcal{C}|_{C_i}$ ($i\in [t-1]$) is a local protection code, and it is an MDS code with parameters $[n_i\leq r+\delta-1,k_i,\delta]$;
\item [(2)] Rank-Size Equtions: For all $i\in [t-1]$,
$$1\leq \rank(\cup_{j=1}^i C_j)-\rank(\cup_{j=1}^{i-1} C_j)=|\cup_{j=1}^i C_j|-|\cup_{j=1}^{i-1} C_j|-(\delta-1),$$
where $C_0:=\emptyset$;
\item [(3)]
$$\rank(\cup_{j=1}^{t-1} C_j\cup \{(j_1,\mathbf{c}_{j_1}),...,(j_i,\mathbf{c}_{j_{s_t}})\})=\rank(\cup_{j=1}^{t-1} C_j)+s_t=k-1;$$
\item [(4)] $U\cap(C_1\cup...\cup C_{t-1}\cup \{(j_1,\mathbf{c}_{j_1}),...,(j_i,\mathbf{c}_{j_{s_t}})\})=\emptyset$ and $|U|=d(\mathcal{C})$.
    \end{itemize}

\underline{Case (II)}:
$$C=C_1\cup ...\cup C_{t} \cup U,$$
where $t=\lceil \frac{k}{r}\rceil -1$, and
\begin{itemize}
\item [(1)] Each $\mathcal{C}|_{C_i}$ ($i\in [t]$) is a protected code, and it is an MDS code with parameters $[n_i\leq r+\delta-1,k_i,\delta]$;
\item [(2)] For all $i\in [t]$,
$$1\leq \rank(\cup_{j=1}^i C_j)-\rank(\cup_{j=1}^{i-1} C_j)=|\cup_{j=1}^i C_j|-|\cup_{j=1}^{i-1} C_j|-(\delta-1),$$
where $C_0:=\emptyset$;
\item [(3)] $\rank(C_1\cup ...\cup C_{t})=k-1$;
\item [(4)] $U\cap (\cup_{j=1}^{t} C_j)=\emptyset$ and $|U|=d(\mathcal{C})$.
\end{itemize}
Conversely, assume that $\mathcal{C}$ is a code with parameters $[n,k,d]$ that exhibits the decomposition of Case (I). If for every $(i,\mathbf{c}_{i})\in \{(j_1,\mathbf{c}_{j_1}),...,(j_{s_t},\mathbf{c}_{j_{s_t}})\}\cup U$, $\mathcal{C}$ possesses $(r,\delta)$-locality for $i$th symbol, then $\mathcal{C}$ is an optimal $(r,\delta)$-LRC. Similarly, if $\mathcal{C}$ has a decomposition of Case (II) and for every $(i,\mathbf{c}_{i})\in U$, $\mathcal{C}$ possesses $(r,\delta)$-locality for $i$th symbol, then $\mathcal{C}$ is also an optimal $(r,\delta)$-LRC.
\end{theorem}

Using the decomposition theorem, we obtain the following technical lemma.

\begin{lemma}\label{lem-tech-x1}
Let $\mathcal{C}$ be an optimal $(r,\delta)$-LRC with parameters $[n,k,d]_q$. Then, there always exist $t=\lceil \frac{k}{r}\rceil$ protection codes $\{C_i\}_{i=1}^t$ satisfying $|C_{i+1}\backslash (\cup_{j=1}^{i} C_j)|\geq \delta$ for $i\in [t-1]$ and $|\cup_{i=1}^t C_i|\geq n-d+\delta$.
\end{lemma}
\begin{proof}
Let $C$ be the set defined as in Eq. \eqref{setC}. By Theorem \eqref{Stru}, there are two decomposition cases for the set $C$. If $C$ belongs to the case (I) and it does not belong to the case (II) in Theorem \eqref{Stru}, i.e., $C$ is written by
$$C=C_1\cup...\cup C_{t-1}\cup \{(j_1,\mathbf{c}_{j_1}),...,(j_i,\mathbf{c}_{j_{s_t}})\}\cup U,$$
where $\{C_i\}_{i=1}^t$ and $\{(j_1,\mathbf{c}_{j_1}),...,(j_{s_t},\mathbf{c}_{j_{s_t}})\}\subseteq T$, and the set $U$ are given in Theorem \eqref{Stru}. Let $C_t=T$. Then, if $|C_{t}\backslash (\cup_{j=1}^{t-1} C_j)|\leq \delta$, then we have $\rank(\cup_{i=1}^{t} C_i)=\rank(\cup_{i=1}^{t-1} C_i)$ due to $C_t$ is local protection code. Hence, we obtain $\rank(\cup_{i=1}^{t} C_i)=k-1$ due to Theorem \eqref{Stru}(I)(3), which contradicts with the assumption that $C$ does not belong to case (II) in Theorem \eqref{Stru}. Thus, we obtain $|C_{t}\backslash (\cup_{j=1}^{t-1} C_j)|\geq \delta$. Applying the Rank-Size Equtions in Theorem \eqref{Stru} and using the fact that $C_i$ is local protection code for $2\leq i\leq t-1$. Using the condition (4) of Theorem \eqref{Stru}(I), we obtain that $|\cup_{i=1}^t C_i|\geq n-d+\delta$. Hence, the local protection codes $\{C_i\}_{i=1}^t$ satisfy $|C_{i+1}\backslash (\cup_{j=1}^{i} C_j)|\geq \delta$ for $i\in [t-1]$ and $|\cup_{i=1}^t C_i|\geq n-d+\delta$.

If $C$ belongs to the case (II) of Theorem \eqref{Stru}, i.e., $C$ is written by
$$C=C_1'\cup...\cup C_{t-1}'\cup U'.$$
Noting that $\rank(C)=k$ and $\rank(C_1'\cup ...\cup C_{t}')=(k-1)$, we can find some $C_t'$ satisfying $\rank(\cup_{i=1}^t C_{t}')=k$. Then, we can use similar discussion as before to obtain that $|(\cup_{j=1}^t C_j')\backslash C_i'|\geq \delta$ for $i\in [t]$ and $|\cup_{i=1}^t C_i'|\geq n-d+\delta$. $\hfill\square$
\end{proof}

Next, we give an example for this lemma.
\begin{example}\label{ex-sh4}
Let $\mathcal{C}$ be the code over $\mathbb{F}_{13}$ whose parity-check matrix is given by
\begin{align}
H=\begin{pmatrix}
A& O\\
O& A \\
B_1& B_2
\end{pmatrix},
\end{align}
where the matrices $A$, $B_1$ and $B_2$ are respectively defined by
\begin{align}
A=\begin{pmatrix}
1&1&1&1&0&0\\
0&0&1&1&1&1
\end{pmatrix}, B_1=\begin{pmatrix}
1&2&3&4&5&6
\end{pmatrix},B_2=\begin{pmatrix}
7&8&9&10&11&12
\end{pmatrix}.
\end{align}
Then, $\mathcal{C}$ is an optimal $(3,2)$-LRC with parameters $[12,7,4]_{13}$. Define the following sets
$$C_1:=\{(1,\mathbf{c}_1),(2,\mathbf{c}_2),(3,\mathbf{c}_3),(4,\mathbf{c}_4)\},\;
C_2:=\{(3,\mathbf{c}_3),(4,\mathbf{c}_4),(5,\mathbf{c}_5),(6,\mathbf{c}_6)\}.$$
Let $C_3=\{(7,\mathbf{c}_7),(8,\mathbf{c}_{8}),(9,\mathbf{c}_{9}),(10,\mathbf{c}_{10})\}$.
Then, the set $\{C_i\}_{i=1}^3$ satisfies $|(\cup_{j=1}^3 C_j)\backslash C_i|\geq 2$ for $1\leq i\leq 3$ and $|(\cup_{j=1}^3 C_j)|=10=n-d+\delta=10$, which coincides with Lemma \ref{lem-tech-x1}.
\end{example}

We recall the following result given in \cite[Theorem 3.9]{ZhouCao2025-Opt}.
\begin{theorem}\label{stru-MDS}
Every local protection code of an optimal $(r,\delta)$-LRC is an MDS code with parameters of the form $[n\leq r+\delta-1,n+1-\delta,\delta]_{q}$.
\end{theorem}

\section{Impurity criterion for Euclidean dual-containing or Hermitian dual-containing optimal $(r,\delta)$-LRCs }\label{general}

In this section, we focus on proving a theorem, which gives a simple criterion for the impurity of Euclidean dual-containing (resp. Hermitian dual-containing) optimal $(r,\delta)$-LRCs.

The following lemma is useful for studying the impurity.

\begin{lemma}\label{basic-ineq}
If $\mathcal{C}$ is an $(r,\delta)$-LRC and it is Euclidean dual-containing or Hermitian dual-containing, then $d\leq r+1$.
\end{lemma}
\begin{proof}
By Lemma \ref{local}, there exists a part of parity-check matrix of $\mathcal{C}$ satisfying $\text{supp}(H')\leq r+\delta-1$ and any $\delta-1$ columns of $H'$ are linearly independent. Thus, the code $\mathcal{C}_{H'}$ generated by $H'$ has parameters $[n'\leq r+\delta-1,k'\geq \delta-1,d']$. By the Singleton bound, $d'\leq n'-k'+1\leq r+\delta-1-k'+1\leq (r+\delta-1)-(\delta-1)+1=r+1$. Since $\mathcal{C}$ is Euclidean dual-containing or Hermitian dual-containing, we obtain that $\mathcal{C}_{H'}$ adding some zero parts belongs to $\mathcal{C}$. Thus, $d(\mathcal{C})\leq d(\mathcal{C}_{H'})=d'\leq r+1$.
$\hfill\square$
\end{proof}

Next, we provide the following impurity criterion.
\begin{theorem}\label{general-purity}
Let $\mathcal{C}$ be an optimal $(r,\delta)$-LRC. Suppose that $\mathcal{C}$ is Euclidean dual-containing or Hermitian dual-containing. If $\mathcal{C}$ induces an impure quantum code, then one of following conditions must hold:
\begin{itemize}
\item[(i)] $d=r+1$;

\item[(ii)] $d\neq r+1$ and $t\leq \max\{\frac{2(r-1)}{r+1-d}, \frac{2(r-1)}{r+\delta-d}+1\}$.
\end{itemize}
\end{theorem}

\begin{proof}
Suppose that neither condition (i) nor condition (ii) holds, i.e., we obtain $d\neq r+1$ and $t>\max\{\frac{2(r-1)}{r+1-d}, \frac{2(r-1)}{r+\delta-d}+1\}$. Next, we only need to show that $\mathcal{C}$ induces pure quantum code in this case and thus obtain a contradictory.

If $t=1$, then $\mathcal{C}$ is an MDS code, which implies that $\mathcal{C}$ induces an pure quantum code. Hence, we assume that $t\geq 2$ in the following proof.

By Lemma \ref{basic-ineq}, we obtain $r+1> d$. For the case that $\mathcal{C}$ is Euclidean dual-containing, as Lemma \ref{lem-tech-x1}, there are some protection codes $\{C_i\}_{i=1}^t$ satisfying that $|C_{i+1}\backslash (\cup_{j=1}^i C_i)|\geq \delta$ for $i\in [t-1]$ and $|\cup_{i=1}^t C_i|\geq n-d+\delta$. By Theorem \ref{stru-MDS}, $C_i$ are MDS codes with the same minimum distance $\delta$ for $i\in [t]$. Suppose that $C_i$ has a parity-check matrix given by
\begin{align}
H_1=(A_{1,1}, A_{1,2}), \;H_i=(A_{i,1},A_{i,2},A_{i,3})\ (2\leq i\leq t-1),\;H_t=(A_{t,1}, A_{t,2})
\end{align}
Then, putting these matrices together and assume that $\mathcal{C}$ has a parity-check matrix $H_i$ given by
\begin{align}
H=
\left (
\begin{array}{cccc cccc ccc}
A_{1,1}& A_{1,2} & & && & & &&&\\
& A_{2,1}&A_{2,2} &A_{2,3} && & & &&&\\
& & &A_{3,1} &A_{3,2}& A_{3,3}& & & &&\\
& & & & & \ddots & & & &&\\
&&& && &A_{t-1,1}&A_{t-1,2}&A_{t-1,3}&& \\
&&& && & & &A_{t,1}&A_{t,2}&A_{t,3}\\
B_{1}& B_{2}&B_{3}&\cdots&\cdots&\cdots &\cdots &\cdots& B_{2t}&B_{2t+1}&B_{2t+2}
\end{array}
\right ),\label{uni-pari}
\end{align}
where $B_i$ is a $l \times v_i$ matrix for some $l\in \mathbb{N}$ and $u_i,v_i\in \mathbb{N}^+$. Now, using the fact that $|C_{i+1}\backslash (\cup_{j=1}^i)|\geq \delta$ for $i\in [t-1]$ and $C_i$ are MDS codes, we obtain that the first $t(\delta-1)$ rows of $H$ are linearly independent. Hence, the matrix $B_i$ has $n-k-t(\delta-1)$ row vectors, i.e., $l=n-k-t(\delta-1)$. Since $\mathcal{C}$ is optimal, we obtain $n-k=d-\delta+t(\delta-1)$, i.e., $n-k-t(\delta-1)=d-\delta$. Hence, we obtain $l=d-\delta$.

Let $S_i=C_i\backslash (C_1\cup \ldots C_{i-1}\cup \ldots C_{t})$ and let $s_i=|S_i|$ for $i\in [t]$. Let $T=\cup_{i=1}^t C_i$. Then, $|T|\geq n-d+\delta$ due to above discussion. First, we claim that there is some $s_i$ satisfying $s_i>d+\delta-2$. Suppose that $s_i\leq d+\delta-2$ for all $i\in [t]$. Then, we obtain
\begin{align}
\sum_{i=1}^t s_i\leq t(d+\delta-2). \label{comp-eq-x1}
\end{align}
By another way, we have $\sum_{i=1}^t s_i\geq [|\cup_{i=1}^t C_i|-(\sum_{i=1}^t |C_i|-|\cup_{i=1}^t C_i|)]=2|\cup_{i=1}^t C_i|-\sum_{i=1}^t |C_i|=2|T|-\sum_{i=1}^t |C_i|$. Hence, we obtain $2|T|\leq \sum_{i=1}^t s_i+\sum_{i=1}^t |C_i|$. Combining this with known fact that $|T|\geq n-d+\delta$ and $|C_i|\leq r+\delta-1$ by definition, we obtain
\begin{align}
2(n-d+\delta)\leq t(d+\delta-2)+t(r+\delta-1). \label{comp-eq-x2}
\end{align}
Since $\mathcal{C}$ is an optimal $(r,\delta)$-LRC, we obtain $n-d=k-1+(t-1)(\delta-1)$. Assume $k=tr-u$, where $0\leq u\leq r-1$. Then, we obtain
\begin{align}
n-d+\delta=t(r+\delta-1)-u. \label{optimal-eq-x1}
\end{align}
Combining this with the inequality \eqref{comp-eq-x2}, we obtain
\begin{align}
2[t(r+\delta-1)-u]\leq t(d+\delta-2)+t(r+\delta-1), \label{comp-eq-x3}
\end{align}
and this is equivalent to $t(r+1-d)\leq 2u$. However, this can not happen since $t(r+1-d)>2(r-1)\geq 2u$ using condition (ii). Hence, the claim holds. Then, we can assume the following inequality without loss of generality:
\begin{align}
s_1>d+\delta-2. \label{ideal-inequ-x1}
\end{align}
Next, we claim that there is some $s_i$ satisfying $s_i\geq d-1$, where $2\leq i\leq t$. Suppose that $s_i\leq d-2$ for all $2\leq i\leq t$. Using similar discussions as before, we obtain that $2(n-d+\delta)\leq s_1+(t-1)(d-2)+t(r+\delta-1)$. Using the fact that $s_1\leq r+\delta-1$, we obtain
\begin{align}
2(n-d+\delta)\leq (t-1)(d-2)+(t+1)(r+\delta-1). \label{ideal-inequ-y1}
\end{align}
Combining this with the known fact $(n-d+\delta)=t(r+\delta-1)-u$ by \eqref{optimal-eq-x1}, we obtain
\begin{align}
2[t(r+\delta-1)-u]\leq (t-1)(d-2)+(t+1)(r+\delta-1), \label{ideal-inequ-y1}
\end{align}
which is equivalent to $(t-1)(r+\delta-d)\leq 2u$. However, this can not happen as $(t-1)(r+\delta-d)>2(r-1)\geq 2u$ using condition (ii). Hence, we can assume the following inequality without loss of generality:
\begin{align}
s_2\geq d-1. \label{ideal-inequ-y2}
\end{align}
Since $s_1> d+\delta-2$ by \eqref{ideal-inequ-x1} and any $d-1$ columns of $H$ are linearly independent, we obtain that the code $\mathcal{C}_1$ whose parity-check matrix is given by the following matrix is an MDS code:
\begin{align}
H_1=\begin{pmatrix}
A_{1,1}\\
B_1
\end{pmatrix}. \label{matrix-tempt4-y1}
\end{align}
Note that $d(C)=\delta+l$, we obtain $d(\mathcal{C}_1)=d$. Hence, we can find some codeword $\mathbf{c}\in \mathcal{C}_1$ such that $\text{wt}(\mathbf{c})=d$. By definition, $H_1\mathbf{c}^\top=\mathbf{0}_{d-1}^\top$, where $\mathbf{0}_{d-1}$ is zero row vector with length $d-1$. Define $\tilde{\mathbf{c}}=(\mathbf{c},\mathbf{0}_{v_2},\ldots,\mathbf{0}_{v_{2t+2}})$, where $\mathbf{0}_{v_i}$ is zero row vector with length $v_i$ ($2\leq i\leq 2t+1$). Next, we claim $\tilde{\mathbf{c}}\in \mathcal{C}$. To show this, we only need to verify that $H\tilde{\mathbf{c}}^\top=\mathbf{0}_{n-k}^\top$, where $\mathbf{0}_{n-k}$ is zero row vector with length $n-k$. This equation holds as $H_1\mathbf{c}^\top=\mathbf{0}_{d-1}^\top$. Hence, $\tilde{\mathbf{c}}\in \mathcal{C}$. Next, we prove that $\tilde{\mathbf{c}} \notin \mathcal{C}^{\perp_\text{E}}$. Suppose that $\tilde{\mathbf{c}} \in \mathcal{C}^{\perp_\text{E}}$. Then, we can express the codeword $\tilde{\mathbf{c}}$ as
\begin{align}
\tilde{\mathbf{c}}=\sum_{i=1}^{n-k}\lambda_i \mathbf{n}_i, \label{ideal-inequ-y3}
\end{align}
where $\lambda_i\in \mathbb{F}$ and $\mathbf{n}_i$ is the $i$th row vectors of $H$. Since $s_2\geq d-1$ by \eqref{ideal-inequ-y2} and any $(d-1)$ columns of $H$ are linearly independent, we see that the code generated by the following matrix is an MDS code:
\begin{align}
\begin{pmatrix}
A_{2,2}\\
B_3
\end{pmatrix}, \label{matrix-tempt2-y2}
\end{align}
which implies that all the row vectors of above matrix \eqref{matrix-tempt2-y2} are linearly independent. Suppose that the all the row vectors of above matrix \eqref{matrix-tempt2-y2} are $\{\mathbf{a}_{i}|\;1\leq i\leq (d-1)\}$. Since $\tilde{\mathbf{c}}=(\mathbf{c},\mathbf{0}_{v_2},\ldots,\mathbf{0}_{v_{2t+2}})$ and the equation \eqref{ideal-inequ-y3}, we obtain
\begin{align}
\sum_{i=1}^{\delta-1}\lambda_{\delta-1+i}\mathbf{a}_{i} +\sum_{i=1}^{l}\lambda_{t(\delta-1)+i}\mathbf{a}_{\delta+i}=\mathbf{0}_{s_2}. \label{ideal-inequ-y3}
\end{align}
Combining this with the already known fact that $\{\mathbf{a}_{i}|\;1\leq i\leq d-1\}$ are linearly independent, we obtain $\lambda_{\delta-1+i}=0$ and $\lambda_{t(\delta-1)+i}=0$ for $1\leq i\leq \delta-1$ and $1\leq j\leq l$. By applying the equation \eqref{ideal-inequ-y3} again and note that the code generated by $(A_{i,1},A_{i,2},A_{i,3})$ are MDS codes with the same dimension $\delta-1$ for $2\leq i\leq t$, we obtain that $\lambda_i=0$ for $\delta\leq i\leq n-k$. This implies that $(\mathbf{c},\mathbf{0}_{v_2})$ belongs to the code $\mathcal{D}$ generated by $(A_{1,1},A_{1,2})$. By Theorem \ref{stru-MDS}, the code $\mathcal{D}$ is an MDS code with dimension $\delta-1$. Moreover, $l(\mathcal{D})\geq s_1>d+\delta-2$ by \eqref{ideal-inequ-x1}, where $l(\mathcal{D})$ is the length of $\mathcal{D}$. This implies that $d(\mathcal{D})=l(\mathcal{D})-(\delta-1)+1>d$. However, this contradicts with the fact that $(\mathbf{c},\mathbf{0}_{v_2})\in \mathcal{D}$ with $\text{wt}((\mathbf{c},\mathbf{0}_{v_2}))=d$, and thus the assumption $\tilde{\mathbf{c}} \in \mathcal{C}^{\perp_\text{E}}$ does not hold, i.e., $\tilde{\mathbf{c}} \notin \mathcal{C}^{\perp_\text{E}}$. By the above discussions, we know that $\tilde{\mathbf{c}}\in \mathcal{C}\backslash(\mathcal{C}^{\perp_\text{E}})$ and $\text{wt}(\tilde{\mathbf{c}})=d$, which implies that $\mathcal{C}$ induces a pure quantum code.

Then, for the case that $C$ is Hermitian dual-containing, we can also obtain the same conclusion by using parallel discussion to the Euclidean dual-containing case. $\hfill\square$
\end{proof}

\begin{remark}\label{purity-finite}
From this result, one can see that if $\mathcal{C}$ is an optimal $(r,\delta)$-LRC with minimum distance $d$ and it is Euclidean dual-containing or
Hermitian dual-containing that satisfies
\begin{itemize}
\item [(1)] $d\neq r+1$;

\item [(2)] $n>t_0(r+\delta-1)+d-\delta$, where $t_0=\max\{\frac{2(r-1)}{r+1-d}, \frac{2(r-1)}{r+\delta-d}+1\}$,
\end{itemize}
then $\mathcal{C}$ induces a pure quantum code by Theorem \ref{general-purity}. This can be seen from the following inequalities:
$$t_0(r+\delta-1)+d-\delta<n=k-1+d+(t-1)(\delta-1)\leq tr-1+d+(t-1)(\delta-1)=t(r+\delta-1)+d-\delta,$$
where the second inequality follows from $k\leq tr$. Moreover, this implies $t>t_0=\max\{\frac{2(r-1)}{r+1-d}, \frac{2(r-1)}{r+\delta-d}+1\}$, which violates the condition (ii) of Theorem \ref{general-purity}.

In particular, if $d\neq r+1$ and $\{C_n\}_{n\in \mathbb{N}^+}$ are optimal $(r,\delta)$-LRCs with the same minimal distance $d$ and they are Euclidean dual-containing or Hermitian dual-containing satisfying $\lim_{n \to \infty} l(C_n)=+\infty$ where $l(C_n)$ is the length of $C_n$, then there is some $m_0\in \mathbb{N}^+$ such that $C_n$ induces pure quantum codes for all $n>m_0$.
\end{remark}

The following example shows that the conditions of Theorem \ref{general-purity} can be only lifted in a small range.
\begin{example}\label{ex-little}
Since Theorem \ref{general-purity} is equivalent to stating that if $\mathcal{C}$ is Euclidean dual-containing or Hermitian dual-containing and the following inequalities hold:
\begin{itemize}
\item[(I)] $r+1-d\neq 0$;

\item[(II)] $t>\max\{\frac{2(r-1)}{r+1-d}, \frac{2(r-1)}{r+\delta-d}+1\}$,
\end{itemize}
then $\mathcal{C}$ induces a pure optimal quantum $(r,\delta)$-LRC. We will give an example for illustrating that these conditions can also only lifted in a very small range by this equivalent form.

Let $\eta$ be a primitive $4$th root of $1$ in $\mathbb{F}_{5^2}$. Let $\mathcal{C}$ be the code over $\mathbb{F}_{5^2}$ whose parity-check matrix is given by
\begin{align}
H=\begin{pmatrix}
A& O&O&O\\
O& A&O&O \\
O& O &A&O \\
O& O &O&A \\
B& B &B&B
\end{pmatrix}, \label{temp-pari-z1}
\end{align}
where $A$ and $B$ are respectively given by
\begin{align}
A=\begin{pmatrix}
1&1&1&1&1
\end{pmatrix}, \ \ B=\begin{pmatrix}
0&1 &\eta&\eta^2&\eta^3
\end{pmatrix}.
\end{align}
We obtain that $d(\mathcal{C})=3$ and $\mathcal{C}$ is both Euclidean dual-containing and Hermitian dual-containing. Moreover, $\mathcal{C}$ is a $(4,2)$-LRC by Lemma \ref{local} and it has parameters $[20,15,3]_{16}$, which implies that $\mathcal{C}$ is an optimal $(4,2)$-LRC. Moreover, the codeword $(\eta-1,-\eta,1,O_{17})\in \mathcal{C}\backslash \mathcal{C}^{\perp_\text{E}}$ and $(\eta-1,-\eta,1,O_{17})\in \mathcal{C}\backslash \mathcal{C}^{\perp_\text{H}}$, where $O_{17}$ is zero row vector with length $17$. From this, we obtain that $\mathcal{C}$ induces a pure quantum code. By definition, we have $r+1=5\neq d=3$, and $\lceil\frac{k}{r}\rceil-\max\{\frac{2(r-1)}{r+1-d}, \frac{2(r-1)}{r+\delta-d}+1\}=4-3=1$. Hence, we conclude that the conditions (I)-(II) above can also only lifted in a very small range.

Another example given by \eqref{impure-vo2} shows that there exist impure quantum $(r,\delta)$-LRCs with minimum distance $d$ satisfying $d\neq r+1$ and $\max\{\frac{2(r-1)}{r+1-d}, \frac{2(r-1)}{r+\delta-d}+1\}-\lceil \frac{k}{r}\rceil=6$. Hence, this also shows that the condition (ii) can be lifted in a small range.
\end{example}

We will also use the following result to judge whether a code induces a pure quantum $(r,\delta)$-LRC or not.
\begin{corollary}\label{coro-key1}
Let $C$ be an optimal $(r,\delta)$-LRC. Suppose that $\mathcal{C}$ is Euclidean dual-containing or Hermitian dual-containing, $r\geq 2d$, and $\lceil\frac{k}{r}\rceil >4$, then $\mathcal{C}$ induces a pure optimal quantum $(r,\delta)$-LRC.
\end{corollary}

\begin{proof}
Since $r\geq 2d$, we have $r+1-d\geq d+1$ which violates the condition (i) of Theorem \ref{general-purity} holds. Note that $\lceil\frac{k}{r}\rceil >4$ implies that $\max\{\frac{2(r-1)}{r+1-d}, \frac{2(r-1)}{r+\delta-d}+1\}\leq 4$. Hence, the condition $\lceil\frac{k}{r}\rceil >4$ makes that the condition (ii) of Theorem \ref{general-purity} does not hold. Therefore, we get what we want.
$\hfill\square$
\end{proof}

Next, we give an example to illustrate this corollary.
\begin{example}\label{ex-pure-sh1}
Recall the optimal codes given in \cite[Theorem 5.4]{ZhouCao2025-Opt} as follows. Let $q\geq 7$. Let $s,v,t\in \mathbb{N}^+$ satisfy
\begin{itemize}
\item  $v|(q-1),\;v\geq 6$ and $s\leq \lfloor \frac{v}{2}\rfloor-2$;

\item  $vt\leq q+v-s-3$.
\end{itemize}
Then, there exists an optimal $(q+v-s-1,s+1)$-LRC with parameters $[t(2q+v-2),t(2q+v-2)-2st-1, s+2]_{q^2}$ which induces an optimal pure quantum $(q+v-s-1,s+1)$-LRC with parameters
\begin{align}
[[t(2q+v-2),t(2q+v-2)-4st-2, s+2]]_{q}. \label{quan-2}
\end{align}
In fact, when $t>4$, the purity of above quantum codes can be directly seen from Corollary \ref{coro-key1}. Note that $r=(q+v-s-1)\geq vt+2\geq v+2\geq 2d=2(s+2)$, and $t=\lceil\frac{k}{r}\rceil >4$. Hence, the above induced quantum code is pure by Corollary \ref{coro-key1}.
\end{example}

\section{Framework and approaches to obtain impure quantum $(r,\delta)$-LRCs and two families of impure quantum $(r,\delta)$-LRCs}

\subsection{A Framework to obtain impure quantum $(r,\delta)$-LRCs}

As guided by Theorem \ref{general-purity}, if we desire to obtain impure quantum $(r,\delta)$-LRCs from classical optimal $(r,\delta)$-LRCs through CSS contruction or Hermitian construction, then we need to study the case that $d=r+1$ or the case that $d\neq r+1$ and $t\leq\max\{\frac{2(r-1)}{r+1-d}, \frac{2(r-1)}{r+\delta-d}+1\}$. In this section, we first give a general way to obtain impure quantum $(r,\delta)$-LRCs from optimal $(r,\delta)$-LRCs by considering the case that $d=r+1$. Then, we consider the case that $d\neq r+1$ and $t\leq\max\{\frac{2(r-1)}{r+1-d}, \frac{2(r-1)}{r+\delta-d}+1\}$ and obtain many impure quantum $(r,\delta)$-LRCs.

We first consider the case that $d=r+1$.
\begin{proposition}\label{pro-impure-x1}
Suppose that $\mathcal{C}$ is a code over $\mathbb{F}_q$. Then, $\mathcal{C}$ satisfies the following conditions:
\begin{itemize}
\item [(1)] $\mathcal{C}$ is an optimal $(r,\delta)$-LRC with minimum distance $\leq 2\delta$;

\item [(2)] there are local protection codes $\{C_i\}_{i=1}^t$ satisfying $C=\cup_{i=1}^t C_i$ and $C_i\cap C_j=\emptyset$ where $C$ is given in \eqref{setC}
\end{itemize}
if and only if the code $\mathcal{C}$ has a parity-check matrix of the form
\begin{align}
H=\begin{pmatrix}
A_1& & &\\
& A_2& & \\
& &\ddots & \\
&& &A_{t} \\
B_1&B_2&\ldots &B_t
\end{pmatrix}, \label{pari-gen-z11}
\end{align}
where
\begin{itemize}
\item[(i)] $A_{i}$ is an $(\delta-1)\times (r_i+\delta-1)$ matrix such that the code generated by $A_{i}$ is an MDS code with dimension $\delta-1$, where $1\leq r_i\leq r$;

\item[(ii)] $B_i$ is an $(d-\delta) \times (r_i+\delta-1)$ matrix for some fixed number $d\leq 2\delta$ satisfying the code generated by the matrix $(A_i^\top,B_i^\top)^\top$ is an MDS code with dimension $d-1$, where $i\in [t]$ and $r_i$ is given in (i).
\end{itemize}
\end{proposition}

\begin{proof}
Suppose that $\mathcal{C}$ is a code satisfying the conditions (1)-(2). Since $C=\cup_{i=1}^t C_i$ and $C_i$ is an MDS code by Theorem \ref{stru-MDS}, we can assume that $C_i$ has parity-check matrix $A_i$ and $\mathcal{C}$ has parity-check matrix given by \eqref{pari-gen-z11} through adding matrices $B_i$ satisfying $\rank(H)=n-k$, where $i\in [t]$. By Theorem \ref{stru-MDS}, we can assume that $A_{i}$ is $(\delta-1)\times (r_i+\delta-1)$ matrix, where $r_i\leq r$. Since $n-k=(d-\delta)+t(\delta-1)$, we obtain that $B_i$ has $n-k-t(\delta-1)=d(\mathcal{C})-\delta$ rows. Hence, $B_i$ is $(d(\mathcal{C})-\delta) \times (r_i+\delta-1)$ matrix, where $d(\mathcal{C})\leq 2\delta$. Now, using the fact that $\mathcal{C}$ has minimum distance $d(\mathcal{C})$, we know that any $d(\mathcal{C})-1$ columns of $H$ are linearly independent. Combining this with the fact that the matrix $(A_i^\top,B_i^\top)^\top$ has exactly $d(\mathcal{C})-1$ rows, we obtain that the code generated by the matrix $(A_i^\top,B_i^\top)^\top$ is an MDS code.

Conversely, suppose that the code $\mathcal{C}$ has a parity-check matrix of the form given in Eq. (\ref{pari-gen-z11}) and meets conditions (i)-(ii) above. Since the form of $H$ and the code generated by $A_{i}$ is MDS, we obtain that $\mathcal{C}$ is an $(r,\delta)$-LRC by Lemma \ref{local}. Next, we show that $d(\mathcal{C})=d$. By the Singleton-like bound, we obtain
\begin{align}
n-k+1-d(\mathcal{C})\geq (t-1)(\delta-1). \label{sn-temp-x1}
\end{align}
By using the condition (ii), we know that all the row vectors of $H$ are linearly independent, i.e., $\rank(H)=t(\delta-1)+d-\delta$. Thus, we have
\begin{align}
n-k=t(\delta-1)+d-\delta. \label{sn-temp-x2}
\end{align}
Combining this with above \eqref{sn-temp-x1}, we obtain $d(\mathcal{C})\geq d$. Next, we only need to show that $d(\mathcal{C})\geq d$, i.e., to show that any $d-1$ columns of $H$ are linearly independent. Denote the $i$th column of $H$ by $\mathbf{h}_i$, where $i\in [n]$. Let $T_i=\{\mathbf{h}_j|\;(i-1)(\delta-1)+1\leq j\leq i(\delta-1)\}$, where $i\in [t]$. Now taking any $d-1$ columns of $H$ denoted by $S=\{\mathbf{h}_{i_1},\ldots, \mathbf{h}_{i_r}\}$. If there is only one set $T_{i_0}$ such that $S\cap T_{i_0}\neq \emptyset$, then $\mathbf{h}_{i_1},\ldots, \mathbf{h}_{i_r}$ are linearly independent since the code generated by the matrix $(A_{i_0}^\top,B_{i_0}^\top)^\top$ is an MDS code. If there are only two sets $T_{i_1}, T_{i_2}$ such that $S\cap T_{i_j}\neq \emptyset$ for $j\in \{1,2\}$, then we can always find $T_{i_{j_0}}$ such that $|S\cap T_{i_j}|\leq \delta-1$ due to $d-1< 2\delta$, where $j_0\in \{1,2\}$. For convenience, we assume that $|S\cap T_{i_1}|\leq \delta-1$ without loss of generality. Assume that $\sum_{h_{i_l}\in T_{i_1}} \lambda_{i_l} \mathbf{h}_{i_l}+
\sum_{h_{i_l}\in T_{i_2}} \lambda_{i_l} \mathbf{h}_{i_l}=\mathbf{0}$. Using the fact that the code generated by $A_{i_1}$ is MDS code, we obtain that $\lambda_{i_l}=0$ for all $\mathbf{h}_{i_l}\in T_{i_1}$. Hence, $\sum_{h_{i_l}\in T_{i_2}} \lambda_{i_l} \mathbf{h}_{i_l}=\mathbf{0}$. From this, we also obtain $\lambda_{i_l}=0$ for all $\mathbf{h}_{i_l}\in T_{i_2}$ due to the code generated by the matrix $(A_{i_2}^\top,B_{i_2}^\top)^\top$ is an MDS code. Hence, $\mathbf{h}_{i_1},\ldots, \mathbf{h}_{i_r}$ are also linearly independent. For general case, we can repeat the above process and obtain that $\mathbf{h}_{i_1},\ldots, \mathbf{h}_{i_r}$ are linearly independent, which implies that $d(\mathcal{C})\geq d$. Hence, we have shown $d(\mathcal{C})=d$. Combining this with equation \eqref{sn-temp-x2}, it follows that $\mathcal{C}$ is an optimal $(r,\delta)$-LRC with minimum distance $d$. Moreover, since the form of $H$, we know that the code $\mathcal{C}$ satisfies the condition (2).
$\hfill\square$
\end{proof}

\begin{remark}
The above proposition is more accurate than the result given in \cite[Theorem 4.2]{ZhouCao2025-Opt} due to the concrete characterization of the minimum distance $d$ through the MDS codes given in (ii).
\end{remark}

Next, we give the following theoretic framework to obtain impure quantum $(r,\delta)$-LRCs from optimal $(r,\delta)$-LRCs.
\begin{theorem}\label{impure-iff}
Suppose that $\mathcal{C}$ is an Euclidean dual-containing (resp. Hermitian dual-containing) optimal $(r,\delta)$-LRC with parameters $[n,k,d]_q$ (resp. $[n,k,d]_{q^2}$). If $\mathcal{C}$ satisfies the following conditions:
\begin{itemize}
\item[(i)] $\delta<d< 2\delta$;

\item[(ii)] there are some non-intersection local protection codes $\{C_i\}_{i=1}^t$ satisfying $C=\cup_{i=1}^t C_i$, where $t=\lceil\frac{k}{r}\rceil$,
\end{itemize}
then $\mathcal{C}$ induces an impure quantum $(r,\delta)$-LRC if and only if the following condition holds:
\begin{align}
n=t(d+\delta-2),\;t\geq 3. \label{impure-equu-x1}
\end{align}
In this case, the code $\mathcal{C}$ has parameters
\begin{align}
[t(d+\delta-2),t(d-1)-d+\delta,d]_q\ (\text{resp. } [t(d+\delta-2),t(d-1)-d+\delta,d]_{q^2}), \label{impure-spec-x1}
\end{align}
which induces an impure quantum $(r,\delta)$-LRC with parameters
\begin{align}
[[t(d+\delta-2),(t-2)(d-\delta), \geq 2\delta]]_{q}. \label{impure-spec-x2}
\end{align}
\end{theorem}

\begin{proof}
Since $\mathcal{C}$ satisfies the conditions (i)-(ii) of Proposition \ref{pro-impure-x1}, it has a parity-check matrix form $H$ given by \eqref{pari-gen-z11} satisfying the conditions (i)-(ii). Suppose that $\mathcal{C}$ induces impure quantum code. Note that $t=1$ implies that the code $\mathcal{C}$ is MDS code, which violates impurity. Similarly, if $t=2$, then the dimension of $\mathcal{C}$ is $r_1+r_2-r+\delta-1$. Since $\mathcal{C}$ is Euclidean dual-containing, we obtain $2\dim(\mathcal{C})\geq n=r_1+r_2+2(\delta-1)$, which implies that $r_1=r_2=r$. From this, we obtain $\mathcal{C}=\mathcal{C}^{\perp_{\text{E}}}$, which also violates impurity. Hence, we have $t\geq 3$. Then, we show that $r_i=d-1$ which implies that $n=t(d+\delta-2)$ and thus the equation \eqref{impure-equu-x1} holds, where $r_i$ is defined in condition (i) of Proposition \ref{pro-impure-x1}. Since the code generated by $A_i$ has minimum distance $r_i+1$ and $\mathcal{C}$ is Euclidean dual-containing, we obtain $r_i+1\geq d$ for all $i\in [t]$. Suppose that there exists some $r_{i_0}+1> d$, then we will obtain a contradictory in the following, implying that $r_i+1= d$ for all $i\in [t]$. Denote the code generated whose parity-check matrix given by $(A_{i_0}^\top,B_{i_0}^\top)^\top$ as $\mathcal{D}$, then we obtain $d(\mathcal{D})=d$ due to the condition (ii). Hence, we can find a codeword $\mathbf{c}\in \mathcal{D}$ satisfying $\text{wt}(\mathbf{c})=d$. Define $\widetilde{\mathbf{c}}=(\mathbf{0}_{r_1+\delta-1},\ldots,\mathbf{0}_{r_{i_0-1}+\delta-1},\mathbf{c},\mathbf{0}_{r_{i_0+1}+\delta-1},\ldots,\mathbf{0}_{r_{t}+\delta-1})$, where $\mathbf{0}_i$ means zero row vector with length $i$. By definition, we know that $\widetilde{\mathbf{c}}\in \mathcal{C}$ with $\text{wt}(\widetilde{\mathbf{c}})=d$. Next, we show that $\widetilde{\mathbf{c}}\notin \mathcal{C}^{\perp_{\text{E}}}$. If $\widetilde{\mathbf{c}}\in \mathcal{C}^{\perp_{\text{E}}}$, then we can find $\{\lambda_i\}_{i=1}^{n-k}\subseteq \mathbb{F}$ satisfying
\begin{align}
\tilde{\mathbf{c}_{0}}=\sum_{i=1}^{n-k}\lambda_i \mathbf{n}_i, \label{impure-tmpt-z1}
\end{align}
where $\mathbf{n}_i$ denotes the $i$th row vector of $H$. Since $t\geq 3$, we can find some $u\in [t]\backslash \{i_0\}$. Since the code generated by the matrix $(A_u^\top,B_u^\top)^\top$ is an MDS code and the above equation \eqref{impure-y1}, we obtain $\lambda_i=0$ for $t(\delta-1)+1\leq i\leq n-k$. Combining this with the equation \eqref{impure-tmpt-z1} again, we obtain that $\mathbf{c}_{0}$ belongs to the code generated by $A_{i_0}$ whose minimum distance is $r_{i_1}$, but this can not happen as $r_{i_1}>\text{wt}(\mathbf{c}_{0})=d$. Thus, we obtain $\widetilde{\mathbf{c}}\notin \mathcal{C}^{\perp_{\text{E}}}$. Since we have shown that $\widetilde{\mathbf{c}}\in \mathcal{C}$ with $\text{wt}(\widetilde{\mathbf{c}})=d$, we obtain that $\mathcal{C}$ induces a pure quantum code, which violates impurity.

Now, suppose that the equation \eqref{impure-equu-x1} holds. Using this condition, we obtain that $\mathcal{C}$ has parameters given by \eqref{impure-spec-x1}. Since the above discussion, we obtain $r_i\geq d-1$ for $i\in [t]$, which implies $n\geq t(d+\delta-2)$. Since the assumption that $n\geq t(d+\delta-2)$, we have $r_i= d-1$ for $i\in [t]$. First, we show that any codeword $\mathbf{c}\in \mathcal{C}$ with $\text{wt}(\mathbf{c})=d(\mathcal{C})$ must belongs to $\mathcal{C}^{\perp_{\text{E}}}$, i.e., $\mathcal{C}$ induces impure quantum code. Suppose $\mathbf{c}\in \mathcal{C}$ with $\text{wt}(\mathbf{c})=d(\mathcal{C})=d$. Let $\mathbf{c}=(\mathbf{c}_1,\ldots,\mathbf{c}_{t})$, where $\mathbf{c}_i$ is row vector with length $(d+\delta-2)$ for $i\in [t]$. Since $H\mathbf{c}^\top=\mathbf{0}_{n-k}$, we obtain
\begin{align}
A_i\mathbf{c}_i^\top=\mathbf{0}_{\delta-1},\;i\in [t]. \label{wt-temp-x1}
\end{align}
Noting that the code whose parity-check matrix given by $A_i$ has minimum distance $\delta$, we have $\text{wt}(\mathbf{c}_i)\geq \delta$ if $\mathbf{c}_i\neq \mathbf{0}_{d+\delta-2}$ by \eqref{wt-temp-x1}. Since $\text{wt}(\mathbf{c})=d(\mathcal{C})=d<2\delta$, we know that there is only one $\mathbf{c}_{i_0}\neq \mathbf{0}_{d+\delta-2}$, i.e., we have
\begin{align}
\mathbf{c}=(\mathbf{0}_{d+\delta-2},\ldots,\mathbf{0}_{d+\delta-2},\mathbf{c}_{i_0},
\mathbf{0}_{d+\delta-2},\ldots,\mathbf{0}_{d+\delta-2}). \label{wt-temp-x2}
\end{align}
Next, we claim that $\mathbf{c}_{i_0}$ belongs to the code $\mathcal{D}$ generated by $A_{i_0}$, which implies that $\mathbf{c}\in \mathcal{C}^{\perp_{\text{E}}}$. Since $H\mathbf{c}^\top=\mathbf{0}_{n-k}$ and the equality \eqref{wt-temp-x2}, we obtain
$A_{i_0}\mathbf{c}_{i_0}^\top=\mathbf{0}_{\delta-1}$ and $B_{i_0}\mathbf{c}_{i_0}^\top=\mathbf{0}_{l}$, where $l=d-\delta$. This implies that $\mathbf{c}_{i_0}$ belongs to the code $\mathcal{D}'$ whose parity-check matrix given by $(A_{i_0}^\top,B_{i_0}^\top)^\top$. We observe that all the row vectors of $A_{i_0}$ belongs to the code $\mathcal{D}'$ due to orthogonality and the matrix $A_{i_0}$ has rank $\dim(\mathcal{D}')=\delta-1$. Hence, $\mathcal{D}=\mathcal{D}'$, which implies that the previous claim holds. So $\mathcal{C}$ induces impure quantum code. Next, we show $d(\mathcal{C}\backslash \mathcal{C}^{\perp_{\text{E}}})\geq 2\delta$. Let $\mathbf{c}'=(\mathbf{c}_1',\ldots,\mathbf{c}_{t}')\in \mathcal{C}\backslash \mathcal{C}^{\perp_{\text{E}}}$, where $\mathbf{c}_i'$ has length $d+\delta-2$ for $i\in [t]$. Then, we obtain $\text{wt}(\mathbf{c}')>d(\mathcal{C})>\delta$. Using similar discussion as before, we obtain that $\text{wt}(\mathbf{c}_i')\geq \delta$ and there are at least two $\mathbf{c}_{i_1}', \mathbf{c}_{i_2}'$ such that $\mathbf{c}_{i_1}'\neq 0_{d+\delta-2}$ and $\mathbf{c}_{i_2}'\neq 0_{d+\delta-2}$. Hence, $\text{wt}(\mathbf{c}')\geq 2\delta$. This implies that $d(\mathcal{C}\backslash \mathcal{C}^{\perp_{\text{E}}})\geq 2\delta$. Combining this with the fact that the code $\mathcal{C}$ has parameters given by \eqref{impure-spec-x1}, we know that $\mathcal{C}$ induces an impure quantum $(r,\delta)$-LRC with parameters given by \eqref{impure-spec-x2}.
$\hfill\square$
\end{proof}

To make the above theorem more computable, we give the following concrete way to obtain impure quantum $(r,\delta)$-LRCs from optimal $(r,\delta)$-LRCs, which depends on the parity-check matrix.

\begin{theorem}\label{impure-fram}
Given positive numbers $r,\delta\geq 2$ satisfying $\delta< r+1< 2\delta$. Suppose that $\mathcal{C}$ is a code. Then, $\mathcal{C}$ is an Euclidean dual-containing (resp. Hermitian dual-containing) optimal $(r,\delta)$-LRC with minimum distance $r+1$ satisfying $C=\cup_{i=1}^t C_i$ for some non-intersection local protection codes $\{C_i\}_{i=1}^t$ if and only if $\mathcal{C}$ has a parity-check matrix of the form given by \eqref{pari-gen-z11} and the matrices $A_i$ and $B_i$ satisfy the following conditions:
\begin{itemize}
\item[(i)] $A_{i}$ is an $(\delta-1)\times (r+\delta-1)$ matrix such that the code generated by $A_{i}$ is an MDS code and it is Euclidean self-orthogonal (resp. Hermitian self-orthogonal), where $i\in [t]$ and $t\geq 3$;

\item[(ii)] $B_i$ is an $(r+1-\delta) \times (r+\delta-1)$ matrix such that the code generated by the matrix $(A_i^\top,B_i^\top)^\top$ is an MDS code, where $i\in [t]$;

\item [(iii)] The code generated by $(B_1,\ldots, B_t)$ is Euclidean self-orthogonal (resp. Hermitian self-orthogonal) and $\langle \mathbf{u}_i, \mathbf{v}_j\rangle_\emph{E}=0$ (resp. $\langle \mathbf{u}_i, \mathbf{v}_j\rangle_\emph{H}=0$), where $\mathbf{u}_i$ is any row vector of $A_i$ and $\mathbf{v}_j$ is any row vector of $B_i$, $i\in [t]$.
\end{itemize}
In this case, the code $\mathcal{C}$ has parameters $[t(r+\delta-1),tr-r-1+\delta,r+1]_q$ (resp. $[t(r+\delta-1),tr-r-1+\delta,r+1]_{q^2}$), which induces an impure quantum $(r,\delta)$-LRC with parameters
\begin{align}
[[t(r+\delta-1),(t-2)(r+1-\delta), \geq 2\delta]]_{q}. \label{impure-quan-y1}
\end{align}
\end{theorem}

\begin{proof}
Note that the condition (i) and the condition (iii) makes that $\mathcal{C}$ to be $\mathcal{C}$ is Euclidean dual-containing (resp. Hermitian dual-containing), then if we combine Proposition \ref{pro-impure-x1} with Theorem \ref{impure-iff} by letting $d=r+1$, then we know that the sufficiency holds. Similarly, we obtain that the necessity holds.
$\hfill\square$
\end{proof}

Next, we consider when the above quantum $(r,\delta)$-LRCs have minimum distance $2\delta$.

\begin{corollary}\label{stru-coro2}
Given positive numbers $r,\delta\geq 2, t\geq 3$ satisfying $\delta< r+1< 2\delta$. Suppose that $\mathcal{C}$ has a parity-check matrix of the form $H$ given by \eqref{pari-gen-z11}, and the matrices $A_i$ and $B_i$ not only satisfy the conditions (i)-(iii) of Theorem \ref{impure-fram} but also satisfy that there exists some $1\leq i_0\neq j_0\leq t$ and $\lambda\in \mathbb{F}$ such that $$(A_{i_0}^\top,B_{i_0}^\top)=(A_{j_0}^\top,\lambda B_{j_0}^\top),\;$$
and the code generated by $(B_{i_0},B_{j_0})$ is Euclidean self-orthogonal (resp. Hermitian self-orthogonal), then the code $\mathcal{C}$ is an optimal $(r,\delta)$-LRC with parameters $[t(r+\delta-1),tr-r-1+\delta,r+1]_q$ (resp. $[t(r+\delta-1),tr-r-1+\delta,r+1]_{q^2}$), which induces an impure quantum $(r,\delta)$-LRC with parameters
\begin{align}
[[t(r+\delta-1),(t-2)(r+1-\delta), 2\delta]]_{q}. \label{impure-quan-y2}
\end{align}
\end{corollary}

\begin{proof}
Suppose that there exists some $1\leq i_0\neq j_0\leq t$ satisfying $(A_{i_0}^\top,B_{i_0}^\top)=(A_{j_0}^\top,\lambda B_{j_0}^\top)$ for some $\lambda\in \mathbb{F}$ and the code generated by $(B_{i_0},B_{j_0})$ is Euclidean dual-containing(resp. Hermitian dual-containing). Note that $B_{i_0}\neq O$ due to the fact that the code $\mathcal{E}$ generated by the matrix $(A_{i_0}^\top,B_{i_0}^\top)^\top$ is an MDS code, where $O$ is zero matrix. Hence, $\lambda\neq 0$. Note that $d(\mathcal{E})=\delta$ due to $\mathcal{E}$ has parity-check matrix $A_{i_0}$, we can find some $\mathbf{c}_{0}\in \mathcal{E}$ such that $\text{wt}(\mathbf{c}_{0})=\delta$. Define $\tilde{\mathbf{c}_{0}}=(\mathbf{c}_{1}'',\ldots,\mathbf{c}_{t}'')$, where $\mathbf{c}_{i_0}''=\mathbf{c}_{0} ,\;\mathbf{c}_{j_0}''=\lambda \mathbf{c}_{0},$ and $\mathbf{c}_{i}''=\mathbf{0}_{r+\delta-1}$ for all $i\in [t]\backslash \{i_0,j_0\}$. Then, by applying that $(A_i^\top,B_i^\top)=(A_j^\top,\lambda B_j^\top)$ for some $\lambda\in \mathbb{F}$ and the code generated by $(B_{i},B_{j})$ is Euclidean dual-containing (resp. Hermitian dual-containing), we have that $H\tilde{\mathbf{c}_{0}}^\top=\mathbf{0}_{n-k}$. Hence, $\tilde{\mathbf{c}_{0}}\in \mathcal{C}$. Suppose that $\tilde{\mathbf{c}_{0}}\in \mathcal{C}^{\perp_{\text{E}}}$. Denote the set of row vectors of $H$ by $\{\mathbf{n}_i|1\leq i\leq (n-k)\}$, where $\mathbf{n}_i$ is $i$th row vector of $H$. Since $\tilde{\mathbf{c}_{0}}\in \mathcal{C}^{\perp_{\text{E}}}$, we can find $\{\lambda_i\}_{i=1}^{n-k}\subseteq \mathbb{F}$ satisfying
\begin{align}
\tilde{\mathbf{c}_{0}}=\sum_{i=1}^{n-k}\lambda_i \mathbf{n}_i. \label{impure-y1}
\end{align}
Since $t\geq 3$, we can find some $u\in [t]\backslash \{i_0,j_0\}$. Since the code generated by the matrix $(A_u^\top,B_u^\top)^\top$ is MDS code and the above equation \eqref{impure-y1}, we obtain $\lambda_i=0$ for $t(\delta-1)+1\leq i\leq (n-k)$. This implies that $\mathbf{c}_{0}$ belongs to the code $\mathcal{D}$ generated by $A_{i_0}$. However, this contradicts with $d(\mathcal{D})=r+1>\text{wt}(\mathbf{c}_{0})=\delta$. Hence, we have shown that $\tilde{\mathbf{c}_0}\in \mathcal{C}\backslash \mathcal{C}^{\perp_{\text{E}}}$ and $\text{wt}(\tilde{\mathbf{c}_0})=2\delta$. Combining this with known fact that $d(\mathcal{C}\backslash \mathcal{C}^{\perp_{\text{E}}})\geq 2\delta$ by Theorem \ref{impure-fram}, we obtain $d(\mathcal{C}\backslash \mathcal{C}^{\perp_{\text{E}}})=2\delta$. Hence, the code $\mathcal{C}$ induces an impure quantum code with parameters given by \eqref{impure-quan-y2}.
$\hfill\square$
\end{proof}

\begin{remark}
The form of parity-check matrices appears in many places (such as \cite{Kong2021New}), and the key to our work lies in identifying the conditions under which the derived quantum codes are impure. This depends on the previous impurity criterion (Theorem \ref{general-purity})  and the characterization of the minimum distance of induced quantum codes. These two aspects are precisely the difficult parts of investigating impure quantum $(r,\delta)$-LRCs.
\end{remark}

Then, we consider the impure case that $d\neq r+1$ and $t\leq\max\{\frac{2(r-1)}{r+1-d}, \frac{2(r-1)}{r+\delta-d}+1\}$ as follows.

\begin{corollary}\label{impure-fram-v2}
Given positive numbers $r,\delta\geq 2,t\geq 3$ satisfying $t<\delta< r+1< 2\delta$. Suppose that the code $\mathcal{C}$ has a parity-check matrix of the form given by \eqref{pari-gen-z11}, then the code $\mathcal{C}$ is an optimal $(r+1,\delta)$-LRC with parameters $[t(r+\delta-1),tr-r-1+\delta,r+1]_q$ (resp. $[t(r+\delta-1),tr-r-1+\delta,r+1]_{q^2}$), which induces an impure quantum $(r+1,\delta)$-LRC with parameters
\begin{align}
[[t(r+\delta-1),(t-2)(r+1-\delta), 2\delta]]_{q}. \label{impure-quan-z1}
\end{align}
\end{corollary}

\begin{proof}
By Corollary \ref{stru-coro2}, the code $\mathcal{C}$ is an optimal $(r,\delta)$-LRC with parameters $[t(r+\delta-1),tr-r-1+\delta,r+1]_q$ (resp. $[t(r+\delta-1),tr-r-1+\delta,r+1]_{q^2}$) which induces an impure quantum $(r,\delta)$-LRC with parameters given in \eqref{impure-quan-z1}. Hence, we only need to show that the code $\mathcal{C}$ is also an optimal $(r+1,\delta)$-LRC. Since $\lceil\frac{\dim(\mathcal{C})}{r+1}\rceil=\lceil\frac{t(r+1)-(r+1+t-\delta)}{r+1}\rceil$
and note that $t<\delta$ and $\delta<r+1$, we obtain that $\lceil\frac{\dim(\mathcal{C})}{r+1}\rceil=t$. This implies that the code $\mathcal{C}$ is also optimal $(r+1,\delta)$-LRC, and therefore we obtain what we want.
$\hfill\square$
\end{proof}

For the code $\mathcal{C}$ in this corollary, note that $r+1-d=1$ and $t\leq\max\{\frac{2(r-1)}{r+1-d}, \frac{2(r-1)}{r+\delta-d}+1\}=2r$, which coincides with the condition (ii) of Theorem \ref{general-purity}.

\subsection{A family of impure quantum $(r,\delta)$-LRCs with unbounded lengths by the CSS construction}
Under the framework of previous Theorem \ref{impure-fram}, we will construct a family of optimal optimal $(r,\delta)$-LRCs with unbounded lengths, which induces impure quantum codes using the CSS construction.

\begin{theorem}\label{impure-inf1}
Assume $q$ is even or $4|(q-1)$. Suppose that $r\geq 1, \delta\geq 2$ and $l\geq 1,t\geq 2$ are positive integers satisfying the following conditions:
\begin{itemize}
\item [(1)] $(r+\delta-1)|(q-1)$;

\item [(2)] $\delta< r+1<2\delta$.
\end{itemize}
Then, there exists an infinite family of optimal $(r,\delta)$-LRCs with parameters $[2t(r+\delta-1),2tr+\delta-r-1,r+1]_q$, which induces impure quantum codes with parameters
\begin{align}
[[2t(r+\delta-1),(2t-2)(r+1-\delta),2\delta]]_q. \label{para-impure-x1}
\end{align}
\end{theorem}

\begin{proof}
Let $m:=(r+\delta-1)$. Since the condition (1), we can find $m$th primitive root of $1$ in $\mathbb{F}_q$, denoted by $\omega$. Define the matrix $A=(a_{i,j})_{1\leq i\leq \delta-1,1\leq j\leq m}$ and the matrix $B=(b_{i,j})_{1\leq i\leq r+1-\delta,1\leq j\leq m}$ over $\mathbb{F}_q$ as follows:
$$a_{i,j}=\omega^{i(j-1)}, \;b_{1,j}=1, \;b_{l,j}=\omega^{(\delta+l-2)(j-1)},\;i\in [\delta-1],\;2\leq l\leq r+1-\delta,\;j\in [m].$$

Since $q$ is even or $4|(q-1)$, we can always find some $\alpha\in \mathbb{F}_q$ satisfying $\alpha^2=-1$. Next, we use Theorem \ref{impure-fram} to obtain impure quantum codes. Using the above matrices, we define the following matrices:
\begin{align}
A_i=A,B_{2j}=\alpha B_{2j-1}=\alpha B, i\in [2t],\; j\in [t]. \label{local-y2}
\end{align}
Let $t\geq 2$ be natural number. Define the code $\mathcal{C}_{2t}$ over $\mathbb{F}_q$ through the following parity-check matrix:
\begin{align}
H=\begin{pmatrix}
A_1& & &\\
& A_2& & \\
& &\ddots & \\
&& &A_{2t} \\
B_1&B_2&\ldots &B_{2t}
\end{pmatrix}. \label{pari-quan-x2}
\end{align}
Next, we check that the parity-check matrix $H$ satisfies the conditions (i)-(iii) of Theorem \ref{impure-fram}, which implies that $\mathcal{C}_{2t}$ is an optimal $(r,\delta)$-LRC with parameters $[2t(r+\delta-1),2tr+\delta-r-1,r+1]_q$ that induces an impure quantum code. The only non-trivial for the above matrix $H$ is the condition (iii) of Theorem \eqref{impure-fram}. To show this, note that the code whose parity-check matrix given by $(B,\alpha B)$ is Euclidean dual-containing by using $\alpha^2=-1$. This implies that the code generated by $(B,\alpha B)$ is Euclidean dual-containing. From this, we know that the code generated by$(B_1,\ldots, B_{2t})$ is Euclidean dual-containing.  Moreover, $\langle \mathbf{u}_i, \mathbf{v}_j\rangle_\emph{E}=0$ due to $\omega$ is primitive $(r+\delta-1)$th root of $1$, where $\mathbf{u}_i$ is any row vector of $A_i$ and $\mathbf{v}_j$ is any row vector of $B_i$, $i\in [2t]$. Hence, the code $\mathcal{C}_{2t}$ is optimal $(r,\delta)$-LRC with parameters $[2t(r+\delta-1),2tr+\delta-r-1,r+1]_q$ which induce impure quantum code by Theorem \eqref{impure-fram}.

To show that $\mathcal{C}_{2t}$ induces an impure quantum code with minimum distance $2\delta$, let $i=1,j=2$ in last part of Theorem \ref{impure-fram}. Then, it satisfies the condition that $(A_1^\top,B_1^\top)=(A_2^\top,\alpha B_2^\top)$, and the code generated by $(B_{1},B_{2})$ is Euclidean dual-containing. Hence, $\mathcal{C}_{2t}$ induces an impure quantum code with parameters given by \eqref{impure-quan-y2} by Corollary \ref{stru-coro2}.
$\hfill\square$
\end{proof}

\begin{remark}\label{exist-x1}
When $q$ is even, then we can let $r=\delta=\frac{q}{2}$ in this theorem, and thus we can obtain infinite families of impure quantum codes with parameters $[[2t(q-1),2t-2,2\delta]]_q$. When $q$ is odd satisfying  $4|(q-1)$, we can let $r=\delta+1=\frac{q-1}{2}$ in this theorem, and therefore we can also obtain infinite families of impure quantum codes with parameters $[[2t(q-1),2t-2,2\delta]]_q$. Hence, for any finite field $\mathbb{F}_q$ satisfying $q$ is even or $4|(q-1)$, the parameters $r$ and $\delta$ in this theorem always exist, which ensures the existence of infinite families of impure quantum codes with unbounded lengths.
\end{remark}

Next, we give an example for the above theorem.
\begin{example}\label{ex-impure-CSS}
Let $q=4$ and let $r=\delta=2$. Then, the numbers $q,r,\delta$ satisfy all the conditions of Theorem \ref{impure-inf1}. Hence, there exists an infinite family of optimal $(2,2)$-LRCs with parameters $[6t,4t-1,3]_4$, which induces impure quantum codes with parameters
\begin{align}
[[6t,2t-2,4]]_4,  \label{para-impure-ex1}
\end{align}
where $t\geq 2$ is a natural number. Observe that the authors in \cite[Remark 8]{Galindo2026} first found an $(2,2)$-LRC with parameters $[[9,1,4]]_3$ induced by optimal $(r,\delta)$-LRC attain the bound in \eqref{eq:delta2-bound}. In fact, this phenomenon is not an isolated phenomenon. When $t=2$, the previous optimal $(2,2)$-LRC induces an impure quantum code with parameters $[[12,2,4]]_4$, which also attains the bound given by \eqref{eq:delta2-bound}.
\end{example}

\subsection{A family of impure quantum codes with unbounded lengths by Hermitian construction}

Let us use the Hermitian construction to obtain an infinite family of impure quantum $(r,\delta)$-LRCs with bounded lengths through optimal $(r,\delta)$-LRCs.
\begin{theorem}\label{impure-inf2}
Suppose that $r\geq 1, \delta\geq 2$ and $t\geq 2,l\geq 1$ are positive numbers satisfying the following conditions:
\begin{itemize}
\item [(1)] $r+\delta-1=q$;

\item [(2)] $\delta< r+1<2\delta$.
\end{itemize}
Then, there exists an infinite family of optimal $(r,\delta)$-LRCs with parameters $[2tq,2tr+\delta-r-1,r+1]_{q^2}$, which induces impure quantum codes with parameters
\begin{align}
[[2tq,(2t-2)(r+1-\delta),2\delta]]_q.  \label{para-impure-x2}
\end{align}
\end{theorem}

\begin{proof}
Let $\omega\in \mathbb{F}_{q^2}$ be a primitive $q-1$th root of $1$. Define the matrix $L=(l_{i,j})_{1\leq i\leq \delta-1,1\leq j\leq q}$ over $\mathbb{F}_{q^2}$ and the matrix $M=(m_{i,j})_{1\leq i\leq r+1-\delta,1\leq j\leq q}$ over $\mathbb{F}_{q^2}$ through defining
$$l_{1,1}=1,\;l_{i,1}=0,\;l_{u,v}=\omega^{(u-1)(v-2)},\;2\leq i\leq \delta-1,\;u\in [\delta-1], \;2\leq v\leq q,$$
and letting
$$m_{s,1}=0,\;m_{s,a}=\omega^{(\delta+s-2)(a-2)},\;s\in [r+1-\delta],\;2\leq a\leq q.$$
Let $u$ be a primitive element of $\mathbb{F}_{q^2}$. Let $\beta\in \mathbb{F}_{q^2}$ satisfying $\beta^{q+1}=-1$ (such as $\beta=u^{\frac{q-1}{2}}$ when $q$ is odd, or $\beta=1$ when $q$ is even). Using the above matrices, we define the following matrices:
\begin{align}
L_i=A,M_{2j}=\beta B_{2j-1}=\beta M, i\in [2t],\; j\in [t]. \label{local-y4}
\end{align}
Let $t\geq 2$ be natural number. We define the code $\mathcal{D}_{2t}$ over $\mathbb{F}_{q^2}$ through the following parity-check matrix:
\begin{align}
H=\begin{pmatrix}
L_1& & &\\
& L_2& & \\
& &\ddots & \\
&& &L_{2t} \\
M_1&M_2&\ldots &M_{2t}
\end{pmatrix}. \label{pari-quan-x3}
\end{align}
Next, we will prove that the code $\mathcal{D}_t$ is an optimal $(r,\delta)$-LRC with parameters $[2tq,2tr+\delta-r-1,r+1]_{q^2}$ that induces an impure quantum code with parameters given by \eqref{para-impure-x2}.

Similar to the proof of Theorem \ref{impure-inf1}, we first verify that the parity-check matrix $H$ satisfies the conditions (i)-(iii) of Theorem \ref{impure-fram}, which implies that $\mathcal{D}_{2t}$ is an optimal $(r,\delta)$-LRC with parameters $[2t(r+\delta-1),2tr+\delta-r-1,r+1]_q$ which induces an impure quantum code. Note that the code whose parity-check matrix given by $(M,\beta M)$ is Hermitian dual-containing by using $\beta^{q+1}=-1$. From this, we know that the code generated by$(M_1,\ldots, M_{2t})$ is Hermitian dual-containing. Moreover, $\langle \mathbf{u}_i, \mathbf{v}_j\rangle_\emph{H}=0$ due to $\omega$ is primitive $q-1$th root of $1$, where $\mathbf{u}_i$ is any row vector of $L_i$ and $\mathbf{v}_j$ is any row vector of $M_i$, $i\in [2t]$. Hence the condition (iii) of Theorem \ref{impure-inf1} holds. Since the condition (i)-(ii) hold due to a direct check, the code $\mathcal{C}_{2t}$ is an optimal $(r,\delta)$-LRC with parameters $[2t(r+\delta-1),2tr+\delta-r-1,r+1]_q$ which induces an impure quantum code by Theorem \eqref{impure-fram}.

To show that $\mathcal{D}_{2t}$ induces an impure quantum code with minimum distance $2\delta$, let $i=1,j=2$ in the final part of Theorem \ref{impure-fram}. Then, it satisfies the condition that $(L_1^\top,M_1^\top)=(L_2^\top,\beta M_2^\top)$ and the code generated by $(M_{1},M_{2})$ is Hermitian dual-containing. Hence $\mathcal{D}_{2t}$ induces an impure quantum codes with parameters given by \eqref{impure-quan-y2} by Theorem \ref{impure-fram}.
$\hfill\square$
\end{proof}

\begin{remark}\label{exist-x2}
When $q>4$ is even, we can set $r=\delta+1=\frac{q}{2}$ in this theorem. Hence we obtain infinite families of impure quantum codes with parameters $[[2tq,4t-4,2\delta]]_q$. When $q$ is odd, we can put $r=\delta=\frac{q+1}{2}$ in this theorem. Hence, we also obtain infinite families of impure quantum codes with parameters $[[2tq,2t-2,2\delta]]_q$. Hence, for any finite field $\mathbb{F}_{q^2}$ satisfying $q>4$ is even or $q$ is odd, the parameters $r$ and $\delta$ in this theorem always exist, and therefore they ensure the existence of infinite families of impure quantum codes with unbounded length.

The reason why we put the parity-check matrix $H$ in the field $\mathbb{F}_{q^2}$ is to use the special property of $\mathbb{F}_{q^2}$ that there is always some $\beta\in \mathbb{F}_{q^2}$ satisfying $\beta^{q+1}=-1$, making the above code $\mathcal{D}_{2t}$ to be Hermitian dual-containing.

Moreover, due to the relation $r+\delta-1=q$ in this theorem for $\mathbb{F}_{q^2}$ and the relation $(r+\delta-1)|(q-1)$ for $\mathbb{F}_{q}$ in Theorem \ref{impure-inf1}, we know that the infinite families of impure quantum codes have different locality from the impure quantum codes given in Theorem \ref{impure-inf1}. Moreover, the length of impure quantum codes in this theorem is fundamentally different from these givne in Theorem \ref{impure-inf1} in a wide range. Hence, these two infinite families of impure quantum codes does not overlap.
\end{remark}

Next, we give an example for the above theorem.
\begin{example}\label{ex-impure-Her1}
Let $q=5$ and $r=\delta=3$. Then, the numbers $q$, $r$, and $\delta$ satisfy all the conditions of Theorem \ref{impure-inf2}. Hence, there exists an infinite family of optimal $(3,3)$-LRCs with parameters $[10t,6t-1,4]_{5^2}$, which induces impure quantum codes with parameters
\begin{align}
[[8t,2t-2,6]]_5,  \label{para-impure-ex2}
\end{align}
where $t\geq 2$ is a positive number. In particular, if $t=2$, then we obtain an optimal $(3,3)$-LRC with parameters $[20,11,4]_{5^2}$, which induces an impure quantum code with parameters $[[20,2,6]]_5$. Its parity-check matrix can be given by
\[
H=
\begin{pmatrix}
L & O & O &O\\
O  & L & O &O\\
O  & O & L &O\\
O  & O & O &L\\
M &  \beta M & M &  \beta M
\end{pmatrix},
\]
where
\begin{align}
L=\begin{pmatrix}
1&1 & 1 &1&1\\
0&1&\omega & \omega^{2}& \omega^{3}
\end{pmatrix} \ \ \mathrm{and}\ \
M=\begin{pmatrix}
0&1& \omega^2& \omega^{4}& \omega^{6}
\end{pmatrix}. \label{local-y4-ex1}
\end{align}
\end{example}

As we have seen that, if $\mathcal{C}$ induces impure quantum codes, then the conditions (i)-(ii) of Theorem \ref{general-purity} must hold. Then, we use Corollary \ref{impure-fram-v2} to obtain impure quantum $(r,\delta)$-LRCs, which coincide with the condition (ii) of Theorem \ref{general-purity}.

\begin{corollary}\label{coro-counter-x1}
Suppose that $r\geq 1,\delta\geq 2$ and $l\geq 1, t\geq 2$ are positive numbers satisfying the following conditions:
\begin{itemize}
\item [(1)] $r+\delta-1=q$;

\item [(2)] $\delta< r+1<2\delta$;

\item [(3)] $2t<\delta$.
\end{itemize}
Then, there exists optimal $(r+1,\delta)$-LRCs with parameters $[2tq,2tr+\delta-r-1,r+1]_{q^2}$, which induces impure quantum codes with parameters
\begin{align}
[[2tq,(2t-2)(r+1-\delta),2\delta]]_q,  \label{impure-coro-x1}
\end{align}
\end{corollary}

\begin{proof}
Combining Theorem \ref{impure-inf2} with Corollary \ref{impure-fram-v2}, we obtain what we want.
$\hfill\square$
\end{proof}

Next, we give an example for this corollary.

\begin{example}\label{impure-vo2}
Let $q,t,r=\delta \geq 2$ be natural numbers satisfying $(2\delta-1)=q$ and $2t<\delta$. Then, the numbers $q,r,\delta,t$ satisfy all the conditions of Corollary \ref{coro-counter-x1}. Hence, there exist optimal $(\delta+1,\delta)$-LRCs with parameters $[2tq,2t\delta-1,\delta+1]_{q^2}$, which induce impure quantum codes with parameters
\begin{align}
[[2tq,2t-2,2\delta]]_{q},  \label{para-impure-ex1}
\end{align}
where $t\geq 2$ is a natural number.

For example, let $q=9,r=\delta=5$ and $t=2$, then there exists an optimal $(6,5)$-LRC with parameters $[36,19,6]_{9^2}$, which induces an impure quantum code with parameters
\begin{align}
[[36,2,10]]_{9}.  \label{para-impure-ex2}
\end{align}
Noting that $\max\{\frac{2(r-1)}{r+1-d}, \frac{2(r-1)}{r+\delta-d}+1\}-2t=10-4=6$, this impure quantum code violates the condition (i) of Theorem \ref{purity-finite} while it obeys the condition (ii) of Theorem \ref{purity-finite}.
\end{example}

\section{Comparison with known impure quantum $(r,\delta)$-LRCs}
Recently, the papers \cite{Galindo2026, Rajpurohit2026} also study impure quantum $(r,\delta)$-LRCs. In \cite{Galindo2026}, the authors use $J$-affine variety codes to obtain impure CSS codes as stated in Theorem 10. Then they give some restriction to the conditions of their \cite[Theorem 10]{Galindo2026} and obtain explicit infinite families of impure CSS codes as stated in \cite[Theorem 12, Theorem 13]{Galindo2026}. Hence, we only need to compare our infinite families of impure quantum $(r,\delta)$-LRCs (Theorems \ref{impure-inf1} and \ref{impure-inf2}) with the impure quantum codes given in \cite[Theorem 10]{Galindo2026}.

Next, we recall that the impure quantum code \(Q(C(\Delta_{H,V,a,b}))\) given in \cite[Theorem 10]{Galindo2026}. It is a quantum \((r,\delta)\)-LRC with parameters
\begin{align}
[[HV,\ 2a+(H \bmod 2),\ (H-\lfloor h+a\rfloor)(v+1)]]_q,  \label{para-compare1}
\end{align}
where \((r,\delta)\)-locality \((v+1,v+1)\). Let $d(H,h,V,v,a)$ be the number defined as
\begin{align}
d(H,h,V,v,a)=\min\{(H-\lfloor h+1\rfloor)(V-\lfloor v\rfloor ),V-\lceil v-1\rceil\}.  \label{para-mini}
\end{align}
Then, we list the impure quantum codes in \cite[Theorem 10]{Galindo2026} as following Table 1.

The authors \cite{Rajpurohit2026} give three families of quantum $(r,\delta)$-LRCs (see \cite[Table 1]{Rajpurohit2026}) and there is one family of quantum $(r,\delta)$-LRCs which produce an impure quantum codes by \cite[Example 2]{Rajpurohit2026}, and we list this family of quantum $(r,\delta)$-LRCs as Table 1.

\begin{table}[htbp]
  \centering
  \caption{Classical and quantum $(r,\delta)$-LRC parameters}
  \resizebox{\textwidth}{!}{%
    \begin{tabular}{@{}llll@{}}
      \toprule
      Reference & Classical $(r,\delta)$-LRC & Quantum $(r,\delta)$-LRC & Restrictions \\
      \midrule
      \cite[Theorem 10]{Galindo2026}
        & \([HV,\,\frac{HV+2a+(H \bmod 2)}{2},\,d(H,h,V,v,a)]_q\)
        & \([[HV,\,2a+(H \bmod 2),\,(H-\lfloor h+a\rfloor)(v+1)]]_q\)
        & \((H-1)\mid(q-1),\ (V-1)\mid(q-1),\ h=\frac{H-1}{2},\  v=\frac{V-1}{2},\ (H-2)\bmod 2\leq a<h\) \\[4pt]
      \cite[Theorem 5]{Rajpurohit2026}
        & \([q-1=gL,\ (g-1)r+(\delta-1),\ r+1]\)
        & \([[gL,\ (g-2)(r-\delta+1),\ \geq r+1]]_q\)
        & \(2 \leq \delta \leq r\) \\[4pt]
        Theorem \ref{impure-inf1}
        & \([2t(r+\delta-1),\,2tr+\delta-r-1,\,r+1]_q\)
        & \([[2t(r+\delta-1),\,(2t-2)(r+1-\delta),\,2\delta]]_q\)
        & \((r+\delta-1)\mid(q-1),\ \delta<(r+1)<2\delta,\ t\ge 2\) \\[4pt]
      Theorem \ref{impure-inf2}
        & \([2tq,\,2tr+\delta-r-1,\,r+1]_{q^2}\)
        & \([[2tq,\,(2t-2)(r+1-\delta),\,2\delta]]_q\)
        & \(r+\delta-1=q,\ \delta<(r+1)<2\delta,\ t\ge 2\) \\[4pt]
      \bottomrule
    \end{tabular}%
  }
\end{table}

First, note that the infinite families of impure quantum codes given in have unbounded length while the impure quantum codes given in \cite[Theorem 10]{Galindo2026} have bounded length due to $HV\leq q^2$. Hence, at least to a large extent, these two families are not overlapping. Second, for the infinite families of impure quantum $(r,\delta)$-LRCs in Theorem \ref{impure-inf1}, their parameters are given in
\begin{align}
[[2t(r+\delta-1),(2t-2)(r+1-\delta),2\delta]]_q.  \label{para-compare2}
\end{align}
where $t\geq 2$ is a natural number and the parameters $r,\delta$ satisfying the conditions (i)-(ii) of Theorem \ref{impure-inf1}. When $r\neq \delta$, the locality $(r,\delta)\neq (v+1,v+1)$, and therefore there is no overlap between the infinite families of impure quantum codes given in Theorem \ref{impure-inf1} and these impure quantum codes with parameters \eqref{para-compare1} in this case. When $r=\delta$, suppose that there is overlap between the parameters given in \eqref{para-compare1} and the parameters given in \eqref{para-compare2}, then one can solve equations and obtain that
$$H=2t,\; V=2\delta-1, \;a=t-1,\;h=t-\frac{1}{2}.$$
Since $(V-1)|(q-1)$ due to the first paragraph of \cite[Section 3]{Galindo2026}, we obtain $(2\delta-2)|(q-1)$ while this can not happen due to $(2\delta-1)|(q-1)$ in condition (i) of Theorem \ref{impure-inf1}. Hence, there is also no overlap between the infinite families of impure quantum codes given in Theorem \ref{impure-inf1} and these impure quantum codes with parameters \eqref{para-compare1} in the case $r=\delta$.

Similarly, for the infinite families of impure quantum codes given in Theorem \ref{impure-inf2}, when $r\neq \delta$, then the locality $(r,\delta)\neq (v+1,v+1)$. Hence, there is no overlap between the infinite families of impure quantum codes given in Theorem \ref{impure-inf2} and these impure quantum codes with parameters \eqref{para-compare1} in this case. When $r=\delta$, suppose that there is overlap between the parameters given by \eqref{para-impure-x2} and the parameters given in \eqref{para-compare2}, then one can also solve equations and obtain that
$$H=2t,\; V=q, \;a=t-1,\;h=t-\frac{1}{2}.$$
Since $(H-1)|(q-1)$ due to the first paragraph of \cite[Section 3]{Galindo2026}, we obtain $(2t-1)|(q-1)$. Hence, when $t>\frac{q}{2}$, then there is also no overlap between the infinite families of impure quantum codes given in Theorem \ref{impure-inf1} and these impure quantum codes with parameters \eqref{para-compare1} in the case $r=\delta$.

Since the minimum distance of their quantum $(r,\delta)$-LRCS \cite[Theorem 5]{Rajpurohit2026} as above are not given explicitly, we can only compare their lower bound case with our minimum distance. Note that the two families of impure quantum codes given in Theorem \ref{impure-inf1} and Theorem \ref{impure-inf2} have minimum distance $2\delta>(r+1)$. Hence, if we let the locality of above quantum $(r,\delta)$-LRCS equal to our locality $(r,\delta)$, we will obtain that the minimum distance of ours is greater than the lower bound of theirs. Note that the authors use Magma software and obtain an impure quantum code with parameters $[[12,2,5]]_{13}$ in \cite[Example 2]{Rajpurohit2026}, and this code is also different from ours. Since our impure quantum codes always have minimum distance $2\delta\neq 5$, there is also no overlap.

Finally, since our classical optimal $(r,\delta)$-LRCs have the property that they are Euclidean dual-containing or Hermitian dual-containing and they induce impure quantum codes, we do not compare classical optimal $(r,\delta)$-LRCs with other classical optimal $(r,\delta)$-LRCs, which do not have the same property.

\section{Concluding remarks}\label{sec:7}

We first establish a criterion, depending only on the parameters \(r, \delta, d, k\), to determine the purity of quantum codes induced from optimal \((r,\delta)\)-LRCs. Consequently, we show that if \(d\neq r+1\), any impure quantum code induced from an optimal \((r,\delta)\)-LRC must have bounded length. On the other hand, we give a framework to obtain impure quantum \((r,\delta)\)-LRCs induced by optimal \((r,\delta)\)-LRCs. As applications, we construct two infinite families of unbounded-length impure quantum codes induced from optimal \((r,\delta)\)-LRCs, all of which satisfy \(d=r+1\). Combining these results, we demonstrate that the condition \(d=r+1\) is a crucial factor controlling the purity of infinite families of unbounded-length quantum codes. Specifically,
\begin{itemize}
  \item If \(d=r+1\), then there exist infinite families of optimal \((r,\delta)\)-LRCs of unbounded
  lengths whose induced quantum codes are impure (Theorem \ref{impure-inf1} and Theorem \ref{impure-inf2}).
  \item If \(d\neq r+1\), then there is no infinite family of optimal \((r,\delta)\)-LRCs of
  unbounded lengths whose induced quantum codes are impure (Theorem \ref{general-purity} and Remark \ref{purity-finite}).
\end{itemize}

The investigation in this paper also prompts the following question for future research:
\begin{itemize}
\item [] Study intersection version of Theorem \ref{impure-fram}, i.e., for an Euclidean dual-containing or Hermitian dual-containing optimal $(r,\delta)$-LRC $\mathcal{C}$, and $C=\cup_{i=1}^m C_i$ for some $m\in \mathbb{N}^+$, where $C$ is defined in \eqref{setC} and $\{C_i\}_{i=1}^m$ are protection codes satisfying that there are some $1\leq i\neq j\leq m$ such that $C_i\cap C_j\neq \emptyset$, then how to use these codes to obtain impure quantum $(r,\delta)$-LRCs?
\end{itemize}

\section*{\large Notes on our discovery process}

Our initial motivation was to provide impurity criteria for quantum \((r,\delta)\)-LRCs derived from optimal \((r,\delta)\)-LRCs, aiming to identify the differences between quantum MDS codes (which are viewed as the quantum $(k,d)$-LRCs with the highest cost of local protection) and ``real'' quantum \((r,\delta)\)-LRCs. Here, ``real'' means $\lceil\frac{k}{r}\rceil>1$.

To find such criteria, a simple and intuitive idea is to find a unified parity-check matrix for these optimal \((r,\delta)\)-LRCs, which is naturally provided by the decomposition theorem (see Theorem \ref{Stru} and the parity-check matrix form given in \eqref{uni-pari} without loss of generality). Using this unified form, we observed that the condition \(d=r+1\) is the key factor influencing the inequality \eqref{comp-eq-x3}.

We then investigated how \(d=r+1\) affects impurity. By exploring the construction of impure quantum \((r,\delta)\)-LRCs for \(d=r+1\), we found that when the protection codes are disjoint, we can characterize the variation range of the minimum distance. This allows us to establish necessary and sufficient conditions for such quantum codes to be impure for $d\leq 2\delta-1$, thereby constructing families of impure quantum \((r,\delta)\)-LRCs with unbounded lengths.

Combining these investigations, we verified that \(d=r+1\) is indeed the key factor controlling the existence of impure quantum \((r,\delta)\)-LRCs with unbounded lengths derived from optimal \((r,\delta)\)-LRCs.

\section*{\large Declaration of competing interest}

The authors declare that they have no known competing financial interests or personal relationships that could have appeared to
influence the work reported in this paper.

\section*{\large Acknowledgment}
K. Zhou is supported by the National Natural Science Foundation of China under Grant No. 12401040.
M. Cao is supported by the National Natural Science Foundation of China under Grant No. 12401684.

\section*{\large Data availability}

No data was used for the research described in this paper.

\end{document}